\documentclass[11pt]{article}
\usepackage{dirtytalk}
\usepackage{amsmath}
\usepackage{graphicx}
\usepackage{amssymb}
\usepackage{gensymb}
\usepackage{float}
\usepackage[style=authoryear,citestyle=authoryear-comp,backend=biber,sorting=nyt,  maxbibnames=99,maxcitenames=2,uniquelist=minyear, dashed=false]{biblatex}
\usepackage{eurosym}
\usepackage{tikz}
\usepackage[raggedright]{titlesec}
\usepackage{caption}
\usepackage{hyperref}
\usepackage[export]{adjustbox}

\usepackage{setspace}   
\newcommand{\seperate}{\noindent\makebox[\linewidth]{\rule{\textwidth}{0.4pt}}

\vspace{1em}}

\numberwithin{equation}{section}

\usepackage[normalem]{ulem}

\begin{document}

\begin{titlepage}
   \begin{center}
       \vspace*{1cm}

       {\huge\textbf{The distribution of aggregate storm risk in a changing climate}}

       \vspace{0.5cm}
       {\Large MTHM005 Mathematical Sciences Project}
            
       \vspace{1.5cm}

       Author - \textbf{Toby P. Jones}\\
       Supervisor - \textbf{David B. Stephenson}

        \vfill
     \singlespacing       
       
{\large\underline{\textbf{Abstract}}}
\begin{flushleft}
The financial losses from extreme weather events can have a disastrous effect, often costing billions of pounds. 
While changes in the disposition of individual events is of importance to both the insurance and re-insurance industries, these companies are often concerned with the aggregate risk posed in a season.  This project explores how the statistical properties of aggregate risk measures may change when, to reflect the earth's changing climate, models are made time dependent. Historical random sum equations by   \textcite{wald} and   \textcite{BandG} are used to develop a relationship between the frequency of events and the aggregate risk.

\vspace{1em}

The covariance between the occurrence of events and aggregate risk is found to be the product of the expected value of the aggregate risk and the dispersion statistic. Furthermore, a new equation (the \say{J-equation}) relates the correlation between the frequency of events and aggregate risk  to the shape of the distribution of storm intensities and the dispersion statistic. This equation highlights that the correlation between the two variables is invariant to a change in the scale of the severity distribution.
The  theory is  applied to a simulated future dataset from the 2020 Norwegian climate model NorESM2-LM. As the data observes the theory presented, this opens the door for these results to be applied to wider geographical regions in future studies.

\end{flushleft}

   \end{center}
\end{titlepage}

\newpage
\tableofcontents
\newpage



\singlespacing
\section{Introduction} \label{Introduction}

While losses incurred from individual storm events can be considerable, such as the
\euro 1.1 billion in estimated losses caused by Storm Ciara in 2020 \parencite{air-alert}, organisations within the insurance and reinsurance industries are not solely concerned with individual events. Instead they are focused on the aggregate risk posed by extreme events, which can be much more severe. \textcite{cusack} highlights the years 1990 and 1999, where windstorms in Europe led to seasonal damages of over \$25 billion. However,   \textcite{hunter_thesis} remarks that in trying to quantify and predict this aggregate risk, insurers use a variety of differing deterministic models. Each model is based upon various assumptions and simplifications of the physical processes involved and subsequently these models often produce contradicting results.   \textcite{hunter} introduced using random sums to quantify the aggregate risk metric, which gives a robust framework to statistically model the yearly risk posed by extremes to overcome the lack of agreement in the insurance industry.

\vspace{1em}

This project aims to build upon the work of \textcite{hunter} and others in the literature, by further exploring the mathematics behind aggregate risk. Many relevant results explored in the literature review in Section \ref{Literature Review} feature in the theory presented in Section \ref{Mathematical modelling}, which also clearly defines aggregate risk and the assumptions behind the process. Section \ref{Mathematical modelling} also revisits accepted results from both   \textcite{wald} and   \textcite{BandG} as well as exploring the relationship between storm frequency and aggregate risk. The application of well-established mathematics on random sums leads to novel equations regarding the correlation between frequency and aggregate risk.  Section \ref{Applications and Results} tests and validates the theory presented on future simulated data from the NorESM2-LM climate model.

\vspace{1em}

New approaches can also be taken from my work, the J-Equation highlights that the correlation between storm frequency and aggregate risk is dependent on the distribution of the severity of storms and the dispersion statistic. \textcite{Economou} notes how the dispersion statistic is expected to increase across much of Europe, the J-Equation implies this will lead to a stronger correlation between storm frequency and aggregate risk.

\newpage
\section{Literature Review } \label{Literature Review}

 Aggregate risk is the cumulative intensity of extratropical cyclones in a season. \textcite{hunter} studied the aggregate risk metric using random sums (as defined in Equation \ref{eqn: Aggregate Risk}), characterising the intensity as the storm's local vorticity. \textcite{hunter} noted how aggregate risk is a suitable measure of the entire value of storm related insurance claims, by assuming the number and size of individual claims are themselves random, the metric captures the seasonal risk posed by cyclones.   \textcite{hunter} also investigated the relationship between the frequency and mean intensity of storms. Using the NCEP-NCAR reanalysis they found a positive correlation over Scandinavia, Northern Germany and the Benelux countries ($\rho =0.2 - 0.6$), with a small negative correlation over the Gulf Stream ($\rho=-0.3$).

\vspace{1em}

  \textcite{hunter} also discussed how negative phases of the Scandinavian Pattern (SCP) are associated with increased cyclone activity, implying it is likely that the SCP accounts for most of the correlation between frequency and intensity. However, it is unknown whether the use of relative vorticity rather than mean sea-level pressure or maximum wind speed would have influenced their results, as this technique is known to detect cyclones earlier in their life. Alongside this, the inclusion of cyclones up to $\pm 5\degree$ away ($\approx 1000$km),  potentially inflates their seasonal count figures.

\vspace{1em}

Whilst teleconnection indices such as the SCP tend to be the main driver of cyclone counts,   \textcite{walz2} points out that studies have linked sea surface temperature (SST) with storm count variability, noting how a horseshoe-shaped anomaly pattern of North Atlantic SST in the Summer and Autumn has a strong link to the North Atlantic Oscillation (NAO) in the subsequent Winter. \textcite{walz2} investigated the variance in storm frequency by exploring links between specific teleconnection indices and the clustering of European storms.

\vspace{1em}

Unlike other studies,   \textcite{walz2} used a wind-based storm tracking algorithm to identify individual events; with their analysis concentrating on the core winter season (December – February). Despite this focus excluding data (storms often occur in October, November and March),  it meant that \textcite{walz2} could better identify more extreme storms (frequently defined as being above the 98th percentile). To quantify the relationship between teleconnection indices and the frequency of events, \textcite{walz2} fitted a Poisson generalised linear model (GLM) to seven slightly overlapping European regions. They then measured cyclone clustering during a season by defining the season to be Active (AS) or Inactive (IAS):
\begin{equation*} \label{Walz Active / Inactive Season Approach}
  \text{Season activity} =
    \begin{cases}
    N_{t,r} > \mu_r + \sigma_r  & \textit{Active} \\
    N_{t,r} < \mu_r + \sigma_r  & \textit{Inactive}. \\
    \end{cases}   
\end{equation*}

  \textcite{walz2} argued this metric was better suited to the actuarial community. For a  year $t$ it clearly communicates storm clustering has occurred in a region $r$ if storm frequency ($N_{t,r}$) is above the level set by long term mean ($\mu_r$) and standard deviation ($\sigma_r$) of storm counts.   \textcite{walz2} observed the highest levels of storm clustering over Scandinavia and, in contrast to thought at the time, concluded the driving factor of storm variability in Europe is the SCP, not the NAO. While these results do not apply to the UK \& Iberian peninsula (whose storms are still directed by the NAO) \textcite{walz2} failed to account for any interaction between the predicting terms (the teleconnection indices) - it is unknown if this would have affected their conclusions.

\vspace{1em}

  \textcite{mailier} introduced the dispersion index (denoted $\phi$ and also called dispersion statistic), with the aim of quantifying the spread of the frequency of storms. \textcite{mailier} defined the index as the ratio of the variance to the mean of $N$:
\begin{equation} \label{Mailer Dispersion Statistic}
    \phi = \frac{ \text{Var}(N) }{ \text{E}[N] } -1.
\end{equation}

A key feature is when the number of storms in a time period follows a Poisson distribution, the dispersion index is zero. When clustering takes place the index is positive, meaning storms are \say{bunched up} with a lack of activity interrupted by multiple storms occurring over a short period in time. Conversely, when the variance in counts is less than the mean ($\phi<0$), this indicates that the distribution of storms in time is a more regular process.

\vspace{1em}

It is known that certain phases of indices such as the NAO and the SCP are positively correlated with European cyclone frequency, as in   \textcite{steptoe}.   \textcite{mailier} also investigated whether teleconnection indices had an impact on the dispersion index. After fitting a Poisson GLM on time-varying teleconnection indices to the mean number of monthly cyclone counts,   \textcite{mailier} found large regions of statistically significant clustering in the European exit region of the North Atlantic storm track. However, the storm tracking algorithm used by   \textcite{mailier} relied on relative vorticity. This resulted in a small southerly shift in the storm tracks ($\approx 1\degree$ / 110km)  when compared to the tracks in similar studies that used Mean Sea Level Pressure (MSLP).

\vspace{1em}

The use of this dispersion index has been challenged by   \textcite{Raschke}, who claimed that the index is inversely proportional to the expected value of storm frequency.   \textcite{Raschke} implies this points to the dispersion of storms being reliant on varying intensity of occurrence in time.   \textcite{Raschke} favours the use of estimating the dispersion parameter by fitting a distribution. They fitted a Generalised Poisson Distribution which is controlled by both a shape and rate parameter to historical storm data in Germany. This enabled them to accurately model both over and under dispersion of counts, a feature which the conventional Poisson and Negative Binomial Distributions are unable to capture. However, this method is computationally expensive, as twice the number of parameters must be estimated when compared to the Poisson distribution.  

\vspace{1em}

 \textcite{Economou} also explored European storm clustering, examining the phenomena in past and future scenario simulations from 17 separate climate models.   \textcite{Economou} found the models were able to capture key geographical features (including regional dispersion and the established European storm tracks) although they concluded that the estimated changes in the NAO are not large enough to create a statistically significant difference to the dispersion of storms.

\vspace{1em}

Despite highlighting the lack of agreement in future model predictions of dispersion (even after accounting for differences in model resolution),   \textcite{Economou} notes clustering is anticipated to increase across the European region. They highlight overdispersion becoming particularly large over Northern Europe, Scandinavia and the Azores. They also found no association  between clustering and model bias, suggesting other processes such as bias in representation of large scale atmospheric variability needs to be taken into account. The only drawback is their use of 30 year datasets, \textcite{cusack} noted how small datasets are vulnerable to sampling variation.

\vspace{1em}

  \textcite{Economou} observed that a changing dispersion statistic could be due to trends in either the mean or variance of cyclone frequency. By calculating modified versions of the dispersion statistic, assuming either mean or variance were invariant to temporal change, they concluded that changes in variation were most significant in driving a shift in dispersion.   \textcite{Economou} concluded that historically, a large element of dispersion can be attributed the changing state of the NAO and its variability. They also outlined how future changes may be driven by the NAO, such as the Atlantic jet stream, which is projected to intensify under future climate conditions. 
  




\vspace{1em}

 \textcite{priestley} aimed to link clustering of European storms to seasonal insurance losses, citing that certain atmospheric dynamics can drive multiple cyclones to focus upon a particular location.     \textcite{priestley} used the high resolution HiGEM climate model to investigate if clustering was more responsible for losses in years when the aggregate risk was more extreme. They used the Generalised Pareto Distribution to model the return period (the average time for an event to occur once) of storms. They concluded that for seasons with return periods greater than three years,  the clustering of cyclones increased the aggregate risk by up to 20\%. 

\vspace{1em}

While \textcite{priestley} has linked the expectation of the aggregate risk to cyclone severity,   \textcite{hunter_thesis} aimed to quantify sources of variation in aggregate risk.   \textcite{hunter} decomposed the sample variance of $S$ as {$s_s^2 = V_n + V_y + V_c$} where $V_n =s_n^2 \bar{y}^2$,
$V_y = s_n^2 \bar{n^2}^2$ and $V_c = \text{cov}(n^2,y^2) - \text{cov}(n,y)^2 - 2\,\text{cov}(n,y) \bar{y}\bar{n}$.

\vspace{1em}

  \textcite{hunter} found the variance of $N$, $V_n$, accounts for 50-80\% of the variance of aggregate loss caused by extra tropical cyclones over the North Atlantic. In this thesis,   \textcite{hunter_thesis} also investigated the dependence between frequency and intensity of natural hazards on aggregate risk, he concluded that assuming independence between frequency and intensity can create large bias in aggregate risk.

\newpage
\section{Mathematical modelling} \label{Mathematical modelling}





\subsection{Introduction}

This section mathematically defines aggregate risk as well as cyclone frequency and intensity and considers some  distribution specific results. The expected value and variance of the aggregate risk is considered before exploring the relationships between aggregate risk, frequency, and intensity. 

\vspace{1em}

Many storms that make landfall to Europe originate in the west, traveling across the Atlantic. Due to the number of factors which can influence the development of a cyclone, the deterministic aspect of the evolution is lost because of uncertainty in the contributing components. As a result, storms may be considered as events that occur at random points in time i.e. as a \textit{point process}. This random occurrence in time means that over a given time period the number (frequency) of storms ($N$) can also be considered to occur at random.

\vspace{1em}

Another aspect of this uncertainty in storm arrival is that the severity of storms may also be treated as occurring at random. We thus obtain a marked point process, in which events arrive at random throughout time each with a random magnitude. For a year $t$, the aggregate risk $S_t$ can therefore be defined as the sum of the intensities $X_{it}$ of the number $N_t$ of storms that arrive:

\begin{equation} \label{eqn: Aggregate Risk}
S_t = X_{1t}+X_{2t}+...+X_{N_{t}t} =\sum_{i=1}^{N_{t}} X_{it}.
\end{equation}

\begin{figure}[ht]
    \centering
    \includegraphics[width=.8\textwidth]{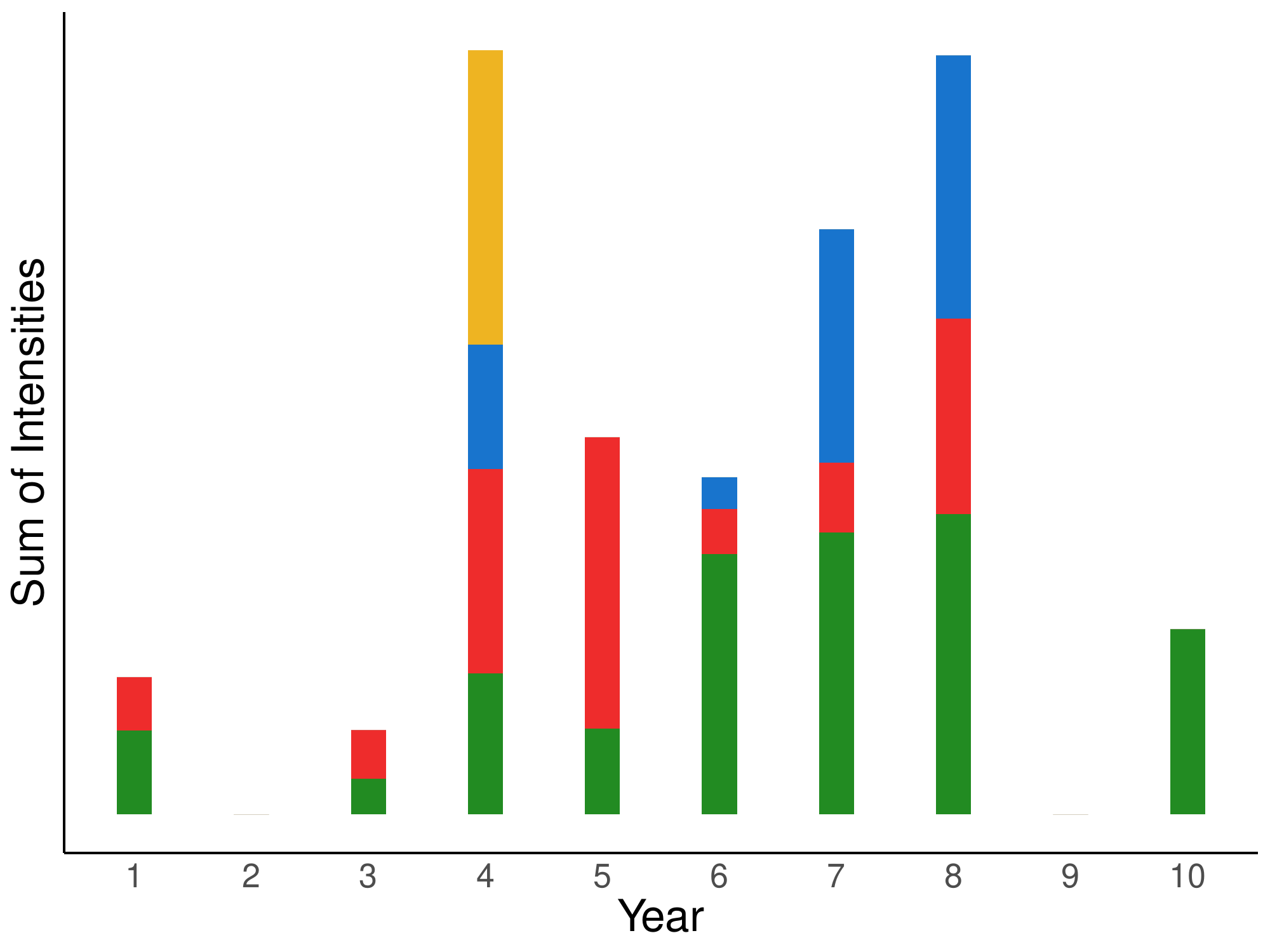}
    \caption{A schematic of the sum of ten Marked Point Processes, with individual occurrences colour coded}
    \label{cumulative marked point process}
\end{figure}

 Equation \ref{eqn: Aggregate Risk} is illustrated in Figure \ref{cumulative marked point process} which shows the motivation behind the study of aggregate risk - years four and eight have similar cumulative values, but with differing event frequency. Previous literature has often assumed temporal stationarity, in that the distributions for $N$ and $X$ do not change with time. This project considers that the frequency and intensity of storms are non-stationary and therefore dependent on time, due to climate change trends.

\vspace{1em}

\vspace{1em}

Figure \ref{fig: S depedennce of N, X and t} displays the relationship between the random variables in the model; for a fixed year $T$, the frequency of storms $N_T$ is independent to each individual storm intensity $X_{iT}$. This relationship is called \say{conditional {independence}} and is denoted $N_T \bot X_{iT}$. The dashed line represents the indirect association that is present due to dependence on $t$. A similar assumption is that for any given year the intensities of storms can also be assumed to be independent and identically distributed (hereafter referred to as \textit{i.i.d}). While   \textcite{hunter}  found correlation between the frequency and intensity of European storms, the values they presented were often small and both positive and negative - therefore for a given year independence over the entire European region is a reasonable assumption to make.

\vspace{1em}

As an example, consider the following more precise mathematical  model. Let $N$ follow a Poisson distribution with rate $\lambda_t$ which itself is exponentially changing in time. Also let the intensities of storms follow a log-Normal distribution with with varying mean $\mu_t$ and variance $\sigma^2$ (note these are not the mean or  variance of the intensities):

\begin{align} \label{eqn: Model Example}
N_t &\sim \text{Poi}(\lambda_t)\nonumber\\
    \lambda_t&=\exp(\alpha_0+\alpha_1 t) \hspace{20pt} \alpha_0, \alpha_1 \in \mathbb{R}\nonumber\\
    \text{log}(X_{it}) &\sim N(\mu_t, \sigma^2) \hspace{20pt} i=1,2,...,N_t\nonumber\\
    \mu_t&=\beta_0+\beta_1 t \hspace{20pt} \beta_0, \beta_1, \sigma^2 \in \mathbb{R}.
\end{align}

While further work in this area could consider a non-constant $\sigma^2$, I keep this stationary. Results for other distributions the storm severity may observe, including the Generalised Pareto Distribution, are shown in Table \ref{table:1} (with my complete derivations in the appendix).



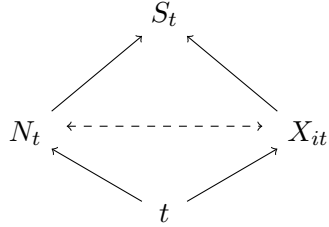
\begin{figure}
\begin{center}
\begin{tikzpicture}
\draw[->] (0,0) -- (1.2,1);
\draw[->] (3,0) -- (1.8,1);
\draw (1.5,1) node[anchor= south] {$S_t$};
\draw (0,0) node[anchor= north east] {$N_t$};
\draw (3,0) node[anchor= north west] {$X_{it}$};
\draw[<-] (0,-0.5) -- (1.2,-1.2);
\draw[<-] (3,-0.5) -- (1.8,-1.2);
\draw (1.5,-1.1) node[anchor= north] {$t$};
\draw[dashed,<->] (0.2,-0.2) -- (2.8,-0.2);
\end{tikzpicture}
\end{center}
  \caption{Illustration of the dependence of Aggregate Risk, $S_t$, on time, $t$}
    \label{fig: S depedennce of N, X and t}
\end{figure}


\subsection{Expected Aggregate Risk}

I will consider the expected risk for a given year: E$[S\,|\,t=T]$\footnote{As \say{conditioning on a fixed year} will continue throughout for ease of notation I will subsequently write
$E_X [X_{iT}\,|\,t=T,N=N_T]$  as $E[X\,|\,T,N].$ }. Using the law of total expectation, it is easy to find E$[S\,|\,T]$:

\begin{align} \label{eqn: Wald Eqn (E[S|T] Proof)}
E[S\,|\,T] &= E\left[\,\sum_{i=1}^{N}X_{iT} \,\bigg |\, T\,\right] \nonumber\\
&= E_N\left[ \,E_X\left[\,\sum_{i=1}^{N} X_{iT} \,\bigg |\, T,N\, \right]\, \right] \nonumber\\
&=E_N\left[ N E [X_{T}\,|\,T,N]\,|\,T\right]\nonumber \\
&=E [N\,|\,T] E [X_{T}\,|\,T].
\end{align}

\vspace{1em}

For a given year, the expected value of the aggregate risk is the product of the expected values of storm frequency and intensity. This equation was presented by   \textcite{wald} and is referred to as \say{Wald's Equation}. 

\newpage
Continuing the precise example I can now calculate the expected value of yearly aggregate risk for a year $t$. The expectation of the Poisson and log-Normal distributions are known results, being  E$[N\,|\,T]=\lambda_T$ and  $\text{E}[X\,|\,T]=\exp\left( \mu_T +\frac{1}{2} \sigma^2\right)$ respectively. As a result, our expected aggregate risk is:
\begin{equation}\label{eqn: Expectation of our model}
    \text{E}[S\,|\,T]=\lambda_T\exp\left( \mu_T +\frac{1}{2} \sigma^2\right).
\end{equation}

\subsection{Variance of Aggregate Risk}

After quantifying the expectation of aggregate risk, in Equation \ref{eqn: Wald Eqn (E[S|T] Proof)}, I aim to find an expression for the variance of $S$ in a given year. I was able to complete this derivation using the \say{Law of Total Variance} which states that for random variables $X$ and $Y$ with Var$(Y)<\infty$ then:
\begin{equation*}\label{eqn: Law of Total Variance}
    \text{Var}(Y)=\text{E}_X\big[\text{Var}_Y(Y\,|\,X)\big]+\text{Var}_X\big(E_Y[Y\,|\,X]\big).
\end{equation*}

By conditioning the variance of $S$ for a given year $T$ on a fixed number of storms $N$ I was able to obtain the result:
\begin{align} \label{eqn: Variance of S Var(S|T)}
\text{Var}\left(S\,|\,T\right)&= \text{E}_N\left[ \text{Var}_X\left(S\,|\,T,N \right) \right] + \text{Var}_N\left( \text{E}_X[S\,|\,T,N] \right) \nonumber\\
&=\text{E}_N\left[ \text{Var}_X\left(\sum_{i=1}^{N} X_{iT}\,\bigg|\,T,N \right) \right] + \text{Var}_N\left( \text{E}_X\left[\sum_{i=1}^{N} X_{iT}\,\bigg|\,T,N\right] \right) \nonumber\\
&= \text{E}\left[ N\,|\,T \right] \text{Var}\left(X\,|\,T \right)  + \text{Var}\left(N\,|\,T\right)\text{E}\left[ X\,|\,T\right]^2  .
\end{align}

This relies on the fact that $X_{iT}$ are \textit{i.i.d} for a given year. This result was first conceived by   \textcite{BandG}, and will particularly be useful later in my exploration of the covariance between storm frequency and aggregate risk. It is worth noting that when $N$ follows a Poisson distribution we have that E$[N\,|\,T]=\text{Var}(N\,|\,T)$ and so Equation \ref{eqn: Variance of S Var(S|T)} can written as:
\begin{align}\label{eqn: Variance of S when N Poisson}
    \text{Var}\left(S\,|\,T\right)&= E\left[ N\,|\,T \right]  E\left[ X^2\,|\,T\right].
\end{align}

\vspace{1em}

As the moments of both the Poisson and log-Normal distribution are well-known, the variance of the precise example in Eqn. \ref{eqn: Model Example} is easy to find:
\begin{align}\label{eqn: Variance of our model}
    \text{Var}\left(S\,|\,T\right)&= E\left[ N\,|\,T \right] E\left[ X^2\,|\,T\right]  \nonumber\\
   &= \lambda_T \exp\left( 2\mu_T +2\sigma^2 \right).
\end{align}

\subsection{The relationship between Frequency and {Aggregate Risk}}

After quantifying the expectation and variance of aggregate risk, it is clear that the frequency and intensity of storms have an impact on aggregate risk. Covariance is a one method of assessing the relationship between two quantities, for two random variables $X$ and $Y$ their covariance is defined as:
\begin{equation}\label{eqn: covariance definition}
    \text{cov}(X,Y)=\text{E}[XY]-\text{E}[X]\text{E}[Y].
\end{equation}
As for a fixed year $N$ and $X$ are conditionally independent, it is clear to see that (from the definition in Equation \ref{eqn: covariance definition})  cov$(N,X\,|\,T)=0$. Therefore we only need to consider cov$(S,X\,|\,T)$ and cov$(S,N\,|\,T)$. In fact, it can be shown that cov$(X,S\,|\,T)=\text{Var}(X\,|\,T)$ (as shown in Derivation \ref{app: Covariance between X and S} in the Appendix). This has straightforward consequences as it implies that the higher the variation in storm intensity, the more influential it is on the value of aggregate risk. Consequently, I will focus on the covariance between storm frequency and aggregate risk:

\begin{align} \label{eqn: Covariance between N and S} 
\text{cov}(N,S\,|\,T)&=E[SN\,|\,T]-E[S\,|\,T]E[N\,|\,T] \nonumber \\
&=E_N[N^2 E_X[X\,|\,T]\,|\,T]- E_N[N\,|\,T]^2 E_X[X\,|\,T] \nonumber\\
&=E_X[X\,|\,T]E_N[N^2 \,|\,T]- E_X[X\,|\,T] E_N[N\,|\,T]^2  \nonumber\\
&=E_X[X\,|\,T]\text{Var}_N(N\,|\,T).
\end{align}


This result implies the covariance between $N$ and $S$ for a given year $T$ is dependent on the variation in the counts and the average intensity. Subsequently, we see the covariance between storm frequency and aggregate risk is $\text{cov}(N,S\,|\,T) =  \phi\text{E}[S\,|\,T]$, more extreme storm seasons mean the relationship between aggregate risk and storm frequency is stronger\footnote{$\phi$ is defined as in Equation \ref{eqn: Dispersion Statistic Def} so is $\phi \ge 0$.}. However, covariance is not always the best measure when comparing the strength of the relationship between sets of two variables, as it is dependent on the magnitude of the variables themselves. A more comparable measure is correlation (denoted $\rho$), which is determined by:

\begin{equation}\label{eqn: PPM Correlation Coefficient}
    \rho =\frac{\text{cov}(X,Y)}{\sqrt{\text{Var}(X)\text{Var}(Y)}}.
\end{equation}

\newpage

Using the definition in Equation \ref{eqn: PPM Correlation Coefficient}, I found an expression for the correlation between $N$ and $S$ for a given year $T$.

\begin{align}\label{eqn: Correlation between N and S} 
\rho&=\frac{\text{cov}(N,S\,|\,T)}{\sqrt{\text{Var}(N\,|\,T)\text{Var}(S\,|\,T)}}      \nonumber \\ 
&=\frac{\text{E}[X\,|\,T]\text{Var}(N\,|\,T)}{\sqrt{\text{Var}(N\,|\,T)\text{Var}(S\,|\,T)}}      \nonumber \\ 
&=\text{E}[X\,|\,T]\sqrt{\frac{\text{Var}(N\,|\,T)}{\text{Var}(S\,|\,T)}}.
\end{align}
Consequently, for the example in Equation \ref{eqn: Model Example}, I found the correlation between $N$ and $S$ for a given year $T$ using Equation \ref{eqn: Correlation between N and S}:
\begin{align}\label{eqn: Model Correlation between N and S}
    \rho&=\text{E}[X\,|\,T]\sqrt{\frac{\text{Var}(N\,|\,T)}{\text{Var}(S\,|\,T)}}      \nonumber \\
    &=\exp \left( \mu_T +\frac{1}{2}\sigma^2 \right)\sqrt{\frac{\lambda_T}{\lambda_T \exp(2\mu_T +2\sigma^2)}}      \nonumber \\
    &=\exp \left(-\frac{1}{2}\sigma^2 \right).
\end{align}

This is an important result, as it suggests  the correlation between frequency and the aggregate risk from storms in a given year is dependent on only the shape parameter of the distribution of storm intensities.

\vspace{1em}

Note that when the frequency of storms follows a Poisson distribution, the influence of frequency of storms on metrics such as correlation tends to \say{drop out} of equations, for example in Equation \ref{eqn: Model Correlation between N and S} there is a cancellation of $\lambda_T$. By introducing a dispersion statistic similar to as defined by   \textcite{mailier}, the expressions for the variance of $S$ as well as covariance (and thus correlation) between $N$ and $S$ can be simplified. The dispersion statistic is defined as:
\begin{equation}\label{eqn: Dispersion Statistic Def}
   \phi = \frac{\text{Var}(N\,|\,T)}{\text{E}[N\,|\,T]}.
\end{equation}
\newpage
Using the expression for the variance (Equation \ref{eqn: Variance of S Var(S|T)}) and the definition shown above it is possible to express the variance of $S$ as a function involving the expected frequency of storms and the dispersion parameter:

\begin{align}\label{eqn: Variance of S in terms of phi}
\text{Var}(S\,|\,T) &= \text{E}[N\,|\,T]\,\text{Var}(X\,|\,T) + \text{Var}(N\,|\,T) \,\text{E}[X\,|\,T]^2  \nonumber \\
&= \text{E}[N\,|\,T]\big( \text{Var}(X\,|\,T) + \phi \text{E}[X\,|\,T]^2  \big) \\
&= \text{E}[N\,|\,T]\big( \text{E}[X^2\,|\,T] + (\phi-1) \text{E}[X\,|\,T]^2  \big)\nonumber. 
\end{align}
As previously mentioned the lack of standardisation means covariance is often difficult to compare. Therefore, an expression for the correlation between $N$ and $S$ allows for ease of comparison between regions and time periods:
\begin{align} \label{eqn: correlation between N and S with phi}
\rho &=\text{E}[X\,|\,T]\sqrt{\frac{\text{Var}(N\,|\,T)}{\text{Var}(S\,|\,T)}}\nonumber\\
 \text{By Eqn \ref{eqn: Variance of S in terms of phi}:}  \hspace{20pt}&=\text{E}[X\,|\,T]\sqrt{\frac{\text{Var}(N\,|\,T)}{\text{E}[N\,|\,T]\big( \text{Var}(X\,|\,T) + \phi \text{E}[X\,|\,T]^2  \big)}}\nonumber\\
 &=\sqrt{\frac{\text{Var}(N\,|\,T)}{\text{E}[N\,|\,T]}}\frac{\text{E}[X\,|\,T]}{\sqrt{\big( \text{Var}(X\,|\,T) + \phi \text{E}[X\,|\,T]^2  \big)}}\nonumber\\
 \text{By Eqn \ref{eqn: Dispersion Statistic Def}:} \hspace{20pt} &=\frac{\sqrt{\phi}\text{E}[X\,|\,T]}{\sqrt{\,\text{Var}(X\,|\,T) + \phi\text{E}[X\,|\,T]^2 \, }}.
\end{align}



When the frequency of storms follows a Poisson distribution, the correlation between $S$ and $N$ is entirely dependent upon the moments of the distribution of storm intensities. I have derived this result for a variety of  distributions the intensity $X_{T}$ may take (the derivations are in the appendix), a summary of these can be seen in Table \ref{table:1}. However, $\phi$ is often not equal to one, as will be seen with the data in Section \ref{Applications and Results}. Another expression which relates $\phi$ to the correlation has a more applications.

From Equation \ref{eqn: correlation between N and S with phi} it follows that: 

\begin{align} \label{eqn: J Equation}
\rho&=\frac{\sqrt{\phi}\text{E}[X\,|\,T]}{\sqrt{\,\text{Var}(X\,|\,T) + \phi\text{E}[X\,|\,T]^2 \, }}\nonumber\\
\implies \rho^2&=\frac{\phi\text{E}[X\,|\,T]^2}{\,\text{Var}(X\,|\,T) + \phi\text{E}[X\,|\,T]^2 \, }\nonumber\\
\implies {\phi\text{E}[X\,|\,T]^2}&=\rho^2(\text{Var}(X\,|\,T) + \phi\text{E}[X\,|\,T]^2 )\nonumber\\
\implies \rho^2\text{Var}(X\,|\,T) &=\phi(1-\rho^2)\text{E}[X\,|\,T]^2\nonumber\\
\frac{\rho^2}{\phi(1-\rho^2)} &=\frac{\text{E}[X\,|\,T]^2}{\text{Var}(X\,|\,T)}\nonumber\\
 \frac{\rho^2}{\phi(1-\rho^2)} &= J^2\\
 &= \left(\frac{\text{E}[X\,|\,T]}{\text{sd}(X\,|\,T)}\right)^2.\nonumber
\end{align}

The variable $J$ is defined as the reciprocal of the coefficient of variation, and I refer to this result (Equation \ref{eqn: J Equation}) as the \textit{J-Equation}. The variable $J$ is dependent only on changes in the shape parameter of the $X_{it}$ distribution, meaning it is invariant to any variation in the  scale parameter. If the dispersion statistic is stationary in time, then any change in correlation must be due to a change in the shape of the distribution of intensities.

\vspace{1em}

Following with the model example, Table \ref{table:1} outlines expressions for the correlation between frequency and intensity as well as the value of the variable $J$. This table also summarises the expected value and variance for when $X_T$ takes differing distributions (assuming $N$ follows a Poisson distribution as before). 

\newpage

\vspace*{\fill}

\begin{table}[H]
\noindent\makebox[\textwidth]{
\begin{tabular}{ |p{2cm}|p{4cm}|p{3.5cm}|p{3.5cm}|p{1.5cm}|p{1.5cm}|  }
\hline
Distribution & $X \sim \cdot$ & E$[S\,|\,T]$ & Var$(S\,|\,T)$ & cor$(N,S)$ & $J^2$\\
\hline
\hline
{Uniform} & ${\text{Unif}(0, \mu_T)}$ & $ \frac{1}{2}\lambda_T\mu_T$ & $\frac{1}{3} \lambda_T \mu_T^2$ & $\frac{\sqrt{3}}{2}$& 3 \\
\hline
{Gamma} & ${\text{Gam}\left(\phi, \frac{1}{\mu_T} \right)}$ & $\phi \lambda_T \mu_T$ & $\lambda_T\mu_T(\phi+\phi^2) ^2$ & $\frac{\phi}{\sqrt{\phi+\phi^2 }}$& $\phi$ \\
\hline
{Exponential} & ${\text{Exp}\left(\frac{1}{\mu_T} \right)}$ & $\lambda_T\mu_T $ & $2\lambda_T\mu_T^2$ & $\frac{\sqrt{2}}{2}$& 1 \\
\hline
{Log-Normal} & ${\text{log}(X) \sim \text{Nor}(\mu_T, \sigma)}$ & $\lambda_T \exp \left( \mu_T + \frac{\sigma^2}{2} \right) $ & $\lambda_T  \exp \left( 2 \mu_T + 2\sigma^2 \right) $& $e^{-\frac{1}{2}\sigma^2}$ & $\frac{1}{ \exp(\sigma^2)-1}$ \\
\hline
{GPD} & ${\text{GPD}\left(0,\frac{1}{\mu_T}, \xi\right)}$ & $ \frac{\lambda_T}{\mu_T(1-\xi)}$ & $\frac{2\lambda_T}{\mu_T^2(1-\xi)(1-2\xi)}$ & $\sqrt{\frac{1-2\xi}{2-2\xi}}$ & $1-2\xi$ \\
\hline
\end{tabular}}
\caption{I have derived these distribution specific examples for the results presented in this section.}
\label{table:1}
\end{table}

\vspace*{\fill}
\newpage 
\section{Application to climate model data} \label{Applications and Results}

\subsection{Introduction}\label{Introduction to the Data and Questions}

\begin{center}
\begin{figure}[H]
    \includegraphics[width=.9\textwidth,center,keepaspectratio]{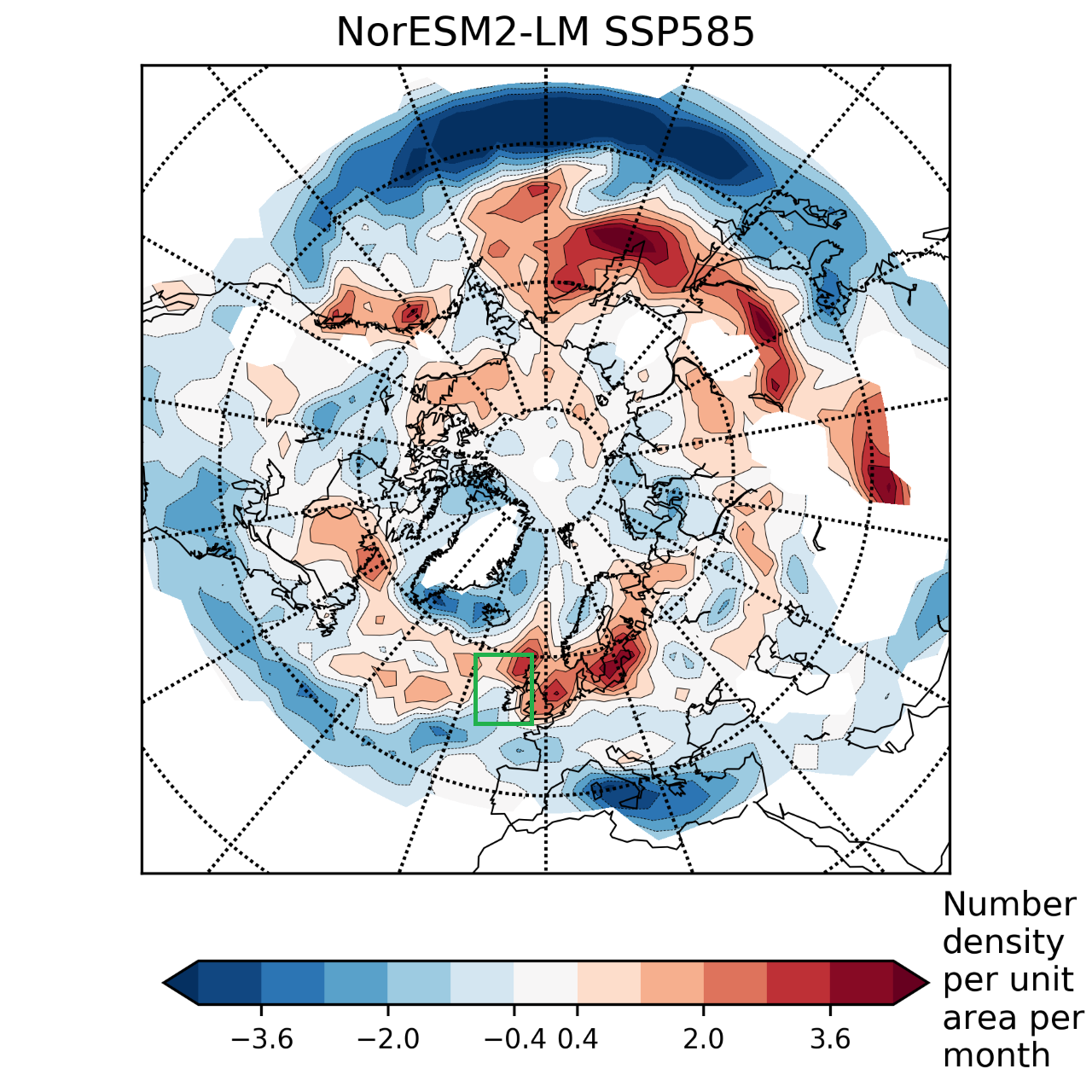}
    \caption{Predicted changes in storm track density - the ratio of mean storm frequency of 2080-2099 compared to the mean frequency of 1979-2014 under the SSP585 climate scenario}
    \label{fig:track density future change map}
\end{figure}    
\end{center}

I have tested the theory by applying it to data from the NorESM2-LM climate model, part of phase 6 of the Coupled Model Intercomparison Project (CMIP6), which contains future predictions of maximum wind speeds from 2040-2099. The particular dataset was chosen as it covers a region to the west of the UK (roughly outlined in green in Figure \ref{fig:track density future change map}), an area which \textcite{zappa} found is likely to have a small ($\approx 3\%$) increase in storm frequency (when comparing mean storm counts from 1976–2005 \&  2070–2099).

Figure \ref{fig:track density future change map}  shows the future predicted changes in storm track density from the NorESM2-LM model under the SSP585 climate scenario. This scenario involves the greatest change to the earth's climate by the end of the century and is the most extreme scenario outlined by \textcite{cmip6_pathways}. As noted by \textcite{priestly2021} this is the largest transformation to storm track density under the climate scenarios considered by CMIP6, this pathway was chosen as it would likely display the strongest signal in my results.  The simulated data was also chosen due to the improved modelling capabilities and increased horizontal resolution CMIP6 models have when compared to their predecessors -   \textcite{priestleydata} notes that the CMIP6 models have less bias present than the older CMIP5 models.

\vspace{1em}

This section addresses the following questions:

\begin{itemize}

\item Does \say{Wald's Equation}  (Equation \ref{eqn: Wald Eqn (E[S|T] Proof)}) hold?

\vspace{0.5em}

\item Does the \say{J-Equation} (Equation  \ref{eqn: J Equation}) hold?

\vspace{0.5em}

\item Can the frequency and intensity of storms in a given year be assumed to be independent?

\vspace{0.5em}

\item How, and to what degree, does the relationship between frequency and aggregate risk change in the future? 

\vspace{0.5em}

\item How well can the occurrence of cyclones in the region be modelled by a Poisson distribution? i.e. Is the dispersion statistic $\phi$ close to one?
\end{itemize}


\vspace{1em}

\begin{figure}[H]
    \centering
    \includegraphics[width=0.8\textwidth,height=\textheight,keepaspectratio]{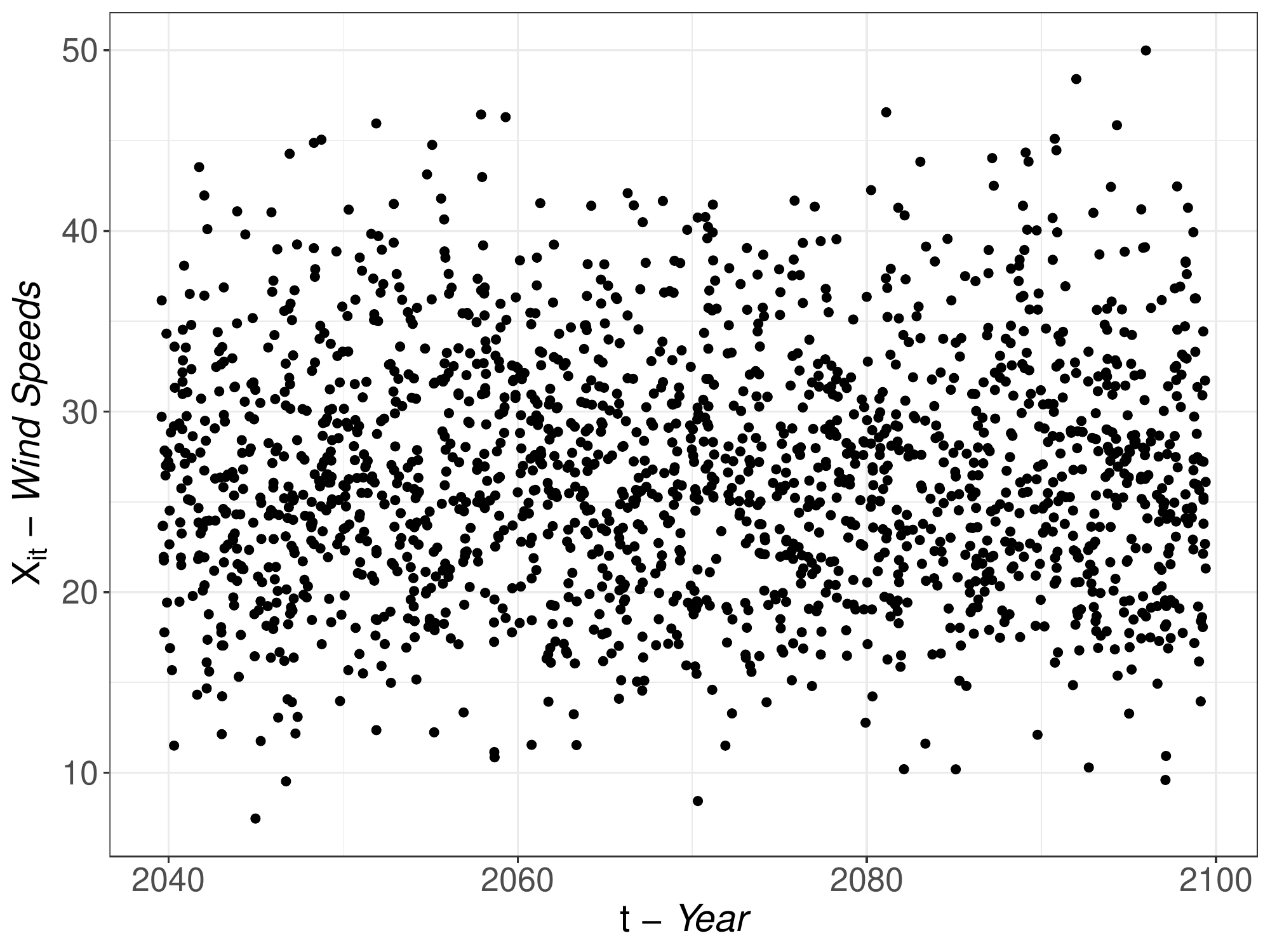}
    \caption{Plot of the raw wind speed values over the duration of the dataset}
    \label{fig: X vs t}
\end{figure}

\begin{figure}[H]
    \centering
    \includegraphics[width=0.8\textwidth,height=\textheight,keepaspectratio]{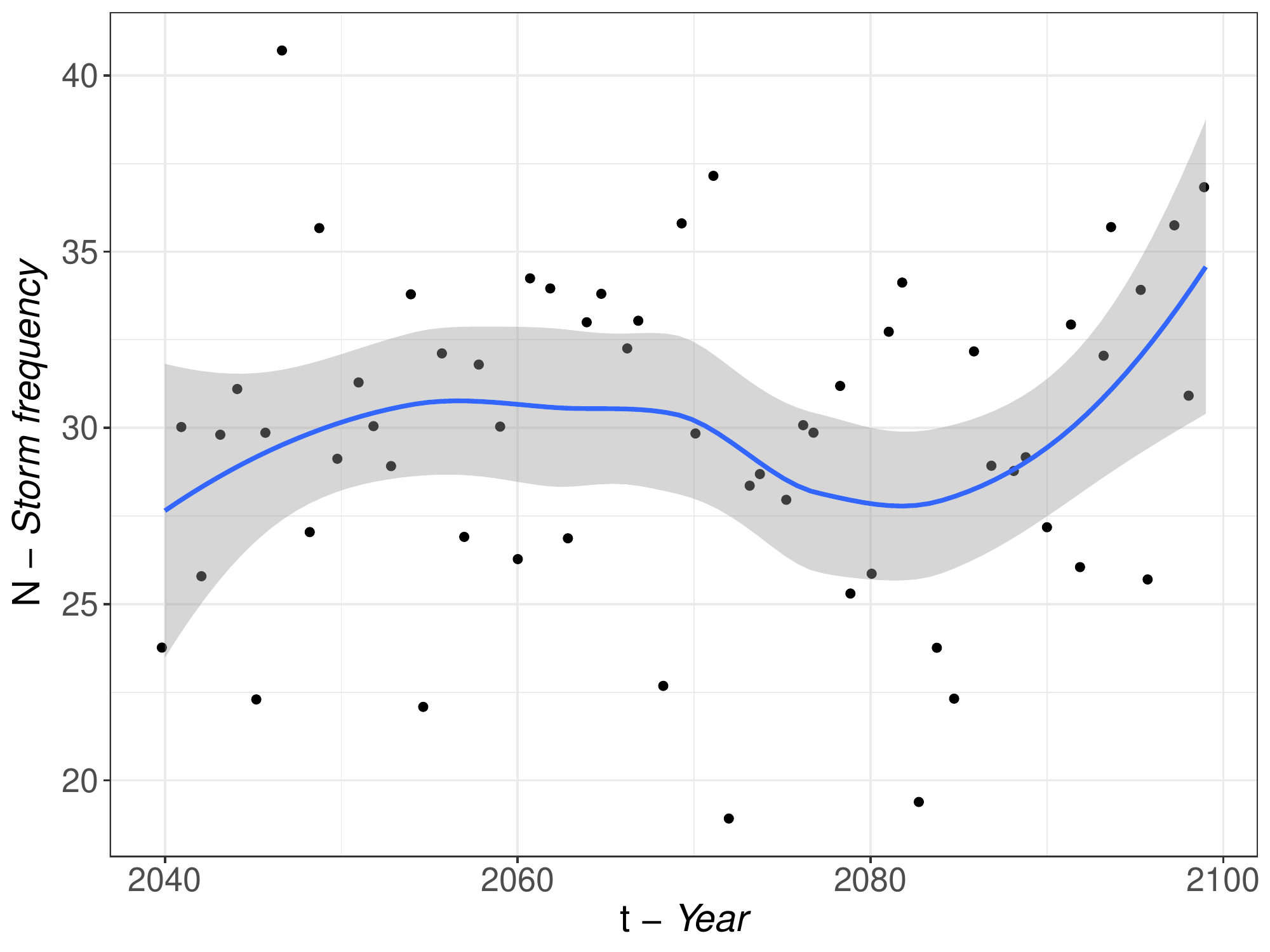}
    \caption{Plot of the number of storms for each year of the dataset}
    \label{fig: N vs t}
\end{figure}

\begin{figure}[H]
    \centering
    \includegraphics[width=0.8\textwidth,height=\textheight,keepaspectratio]{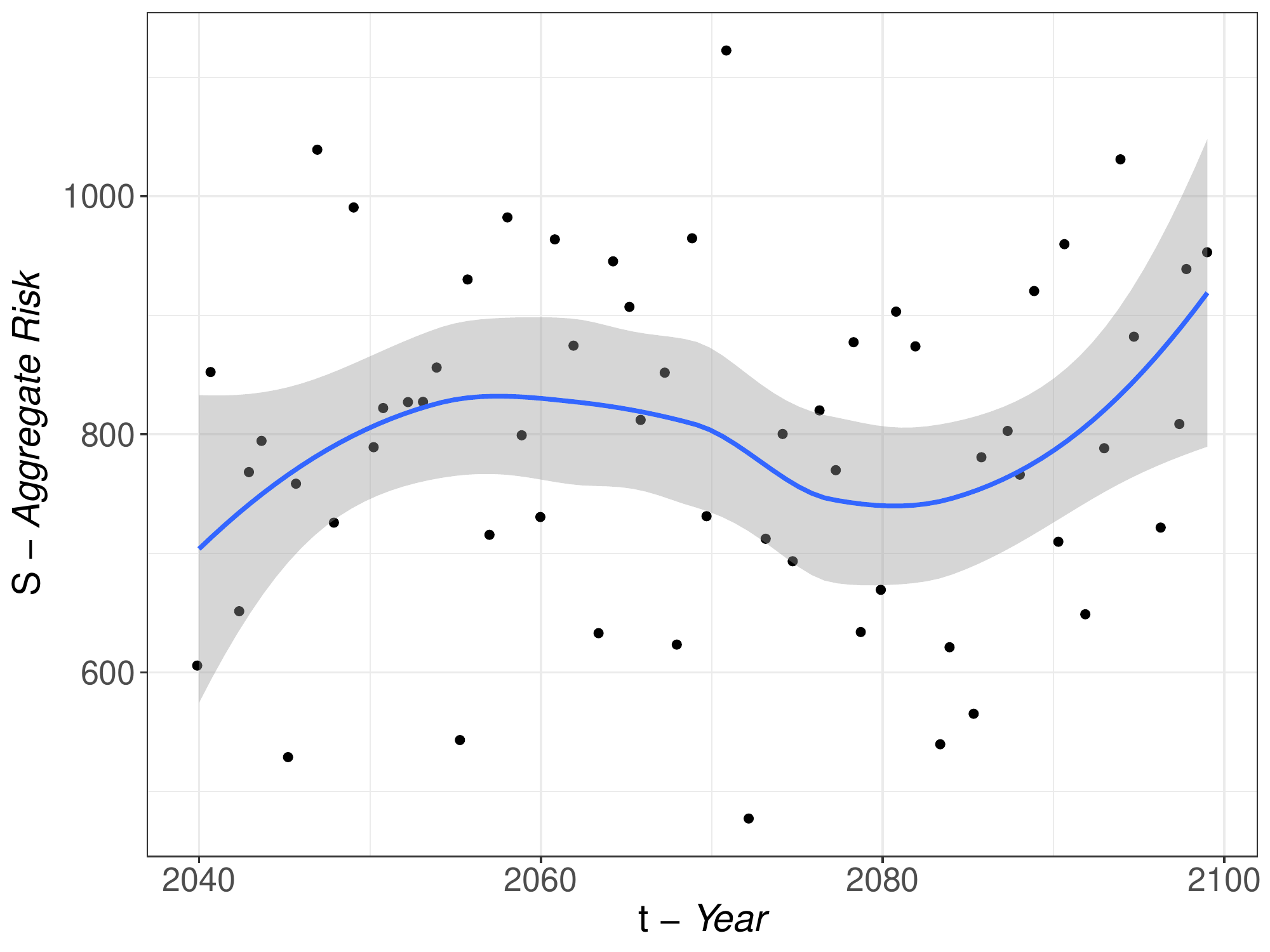}
    \caption{Plot of yearly aggregate risk for the dataset}
    \label{fig: S vs t}
\end{figure}

Figures \ref{fig: X vs t},  \ref{fig: N vs t}, and  \ref{fig: S vs t} depict the storm intensities, storm frequency and seasonal aggregate risk for the dataset. While Figure \ref{fig: X vs t} shows little to no trend in time, with storms being fairly stationary, the values of aggregate risk and storm frequency follow a similar trend for the first 30 years of the dataset. There is a slight decrease around 2070, but an overall increase in frequency and aggregate risk in time.

\subsection{Evaluation of assumptions}\label{Examination into the rigour of Equations}

Wald's Equation (Equation \ref{eqn: Wald Eqn (E[S|T] Proof)}) states the expected value of aggregate risk is equal to to the product of the expected storm frequency ($N$) and expected storm intensity ($X$). Subsequently for a given year the estimate of E$[X]$ would be E$[X]=\frac{1}{N} \sum_i^N X_i = \frac{S}{N}$. As a result, it is clear that  Wald's Equation holds for a fixed year. 

\vspace{1em}

With non-stationary models, finding the \say{true value} of the aggregate risk is difficult as for a given year there is only one observation. Another approach is to find an estimate of the true value over a time period from year 1 to $t$ by using the mean value over that period. As the distributions of storm intensity  are non-stationary, the law of large numbers does not apply. However, Kolmogorov's Strong Law of Large Numbers (see \textcite{Kolmogorov}) can be applied. This states that for independent, but \textit{not} identically distributed, random variables $K_t$ ($t=1, \cdots , T$) that if every $K_t$ has a finite second moment and if $\sum_{t=1}^\infty \frac{1}{t^2} \text{Var}(K_t)<\infty$, then:
\begin{equation} \label{App: Kolmogorov's Strong Law}
    \bar K_t - \text{E}[\bar K_t ] \xrightarrow{\textit{a.s.}}0.
\end{equation}
\newpage
As the earth's climate system is physically unable to generate infinite variance  of both storm counts and severity the law can be applied here, meaning as $t \rightarrow \infty $ the mean of our random variable  converges \textit{almost surely} to the its expected value.  While these equations are only defined for $N_y \ne 0$, the region covered by the dataset is large enough so that $N_y>0$ for all years. I found the long run expected value of $N$, $S$ and $X$ by computing:

\begin{align}\label{N S X Equations}
     \lim_{t \to \infty} \frac{\sum_{y=1}^{t} N_y  }{\sum_{y=1}^{t} 1 }&= \text{E}[N(t)]\nonumber\\
     \lim_{t \to \infty}  \frac{\sum_{y=1}^{t} \sum_{i=1}^{N_y}X_{yi}}{\sum_{y=1}^{t}1} &= \text{E}[S(t)]\nonumber\\
      \lim_{t \to \infty}  \frac{\sum_{y=1}^{t} \sum_{i=1}^{N_y}X_{yi}} {\sum_{y=1}^{t} \sum_{i=1}^{N_y} 1 } &= \text{E}[X(t)].
 \end{align}

\begin{figure}[H]
    \centering
    \includegraphics[height=0.425\textheight,keepaspectratio]{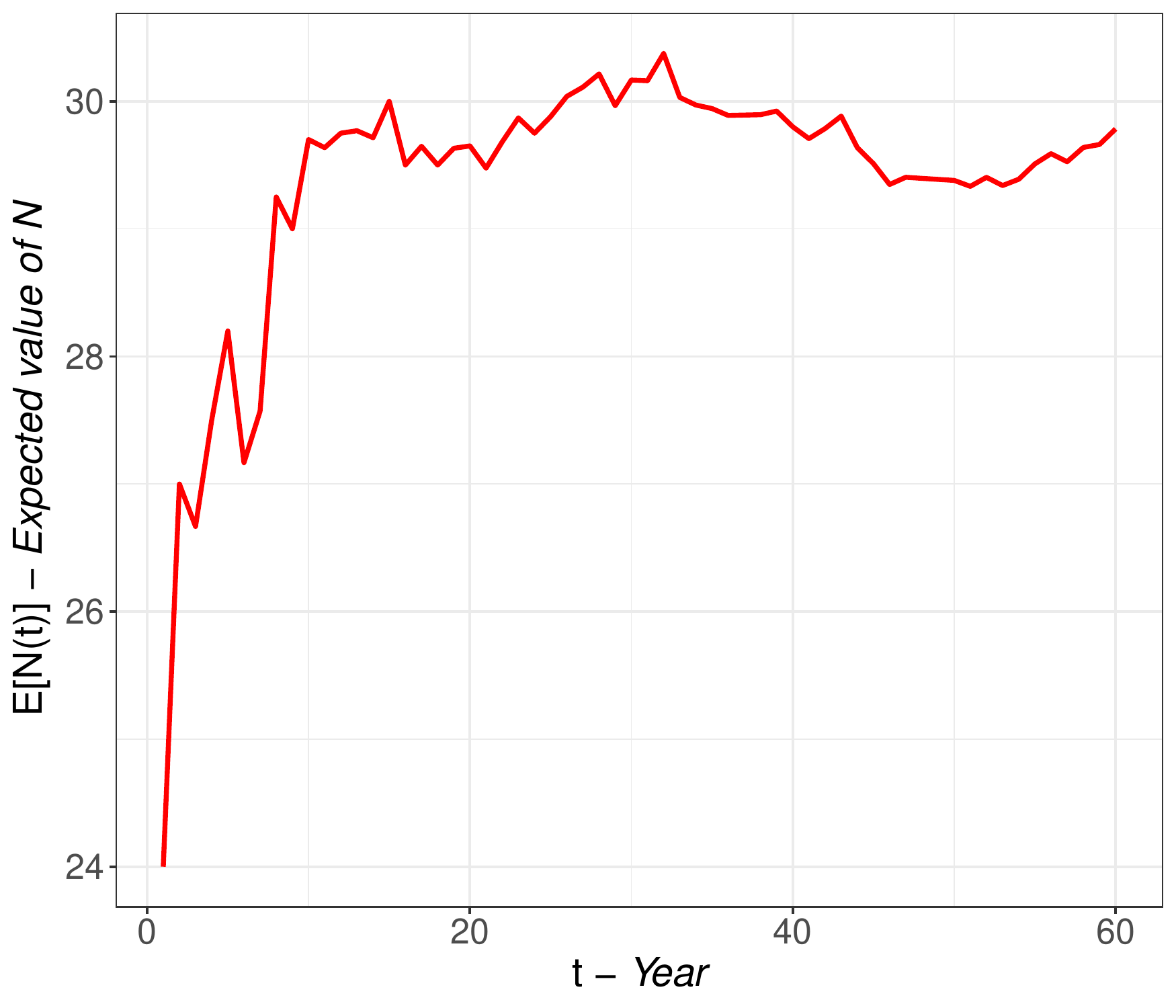}
    \caption{Long run average of $N(t)$ duration of the dataset}
    \label{fig: Long Run N(t)}
\end{figure}

\begin{figure}[H]
    \centering
    \includegraphics[height=0.4\textheight,keepaspectratio]{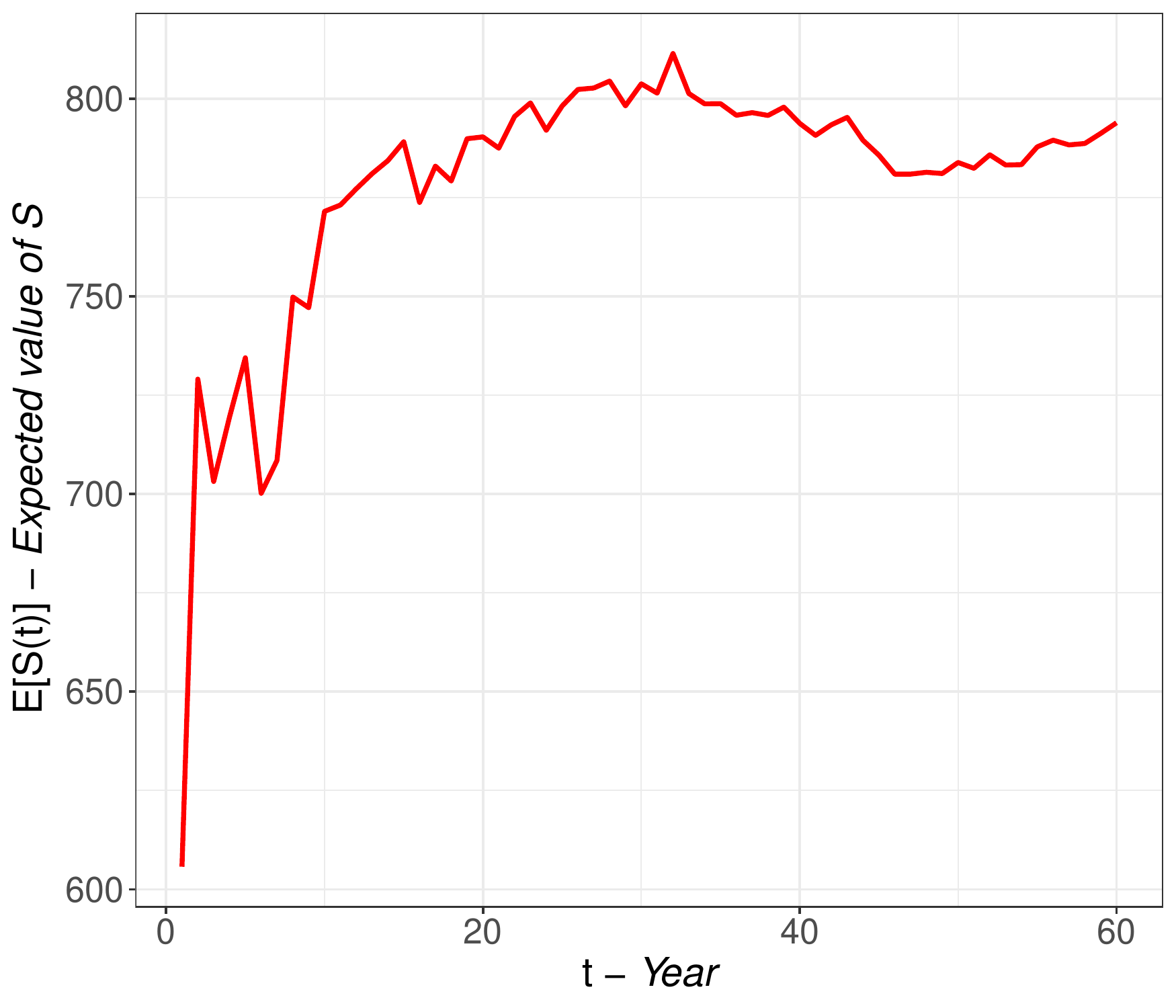}
    \caption{Long run average of $S(t)$ over the duration of the dataset.}
    \label{fig: Long Run S(t)}
\end{figure}

Proof \ref{App: Long Run Wald Derivation} in the Appendix shows that the long-run averages in Eqn. \ref{N S X Equations} do follow Wald's Equation. Figure \ref{fig: Long Run N(t)} demonstrates the convergence of $N$ to its expectation ($\approx 29.5$ per year), while Figure \ref{fig: Long Run S(t)} shows the convergence of $S$ to its long-run expected value ($\approx 785$ms$^{-1}$). Figure \ref{App: Long Run X(t)} (in the appendix) shows the long-run expected value of $X$ converges to $E[X]\approx 26.7$ms$^{-1}$.

\vspace{1em}

\begin{figure}[H]
\centering
\includegraphics[height=0.425\textheight,keepaspectratio]{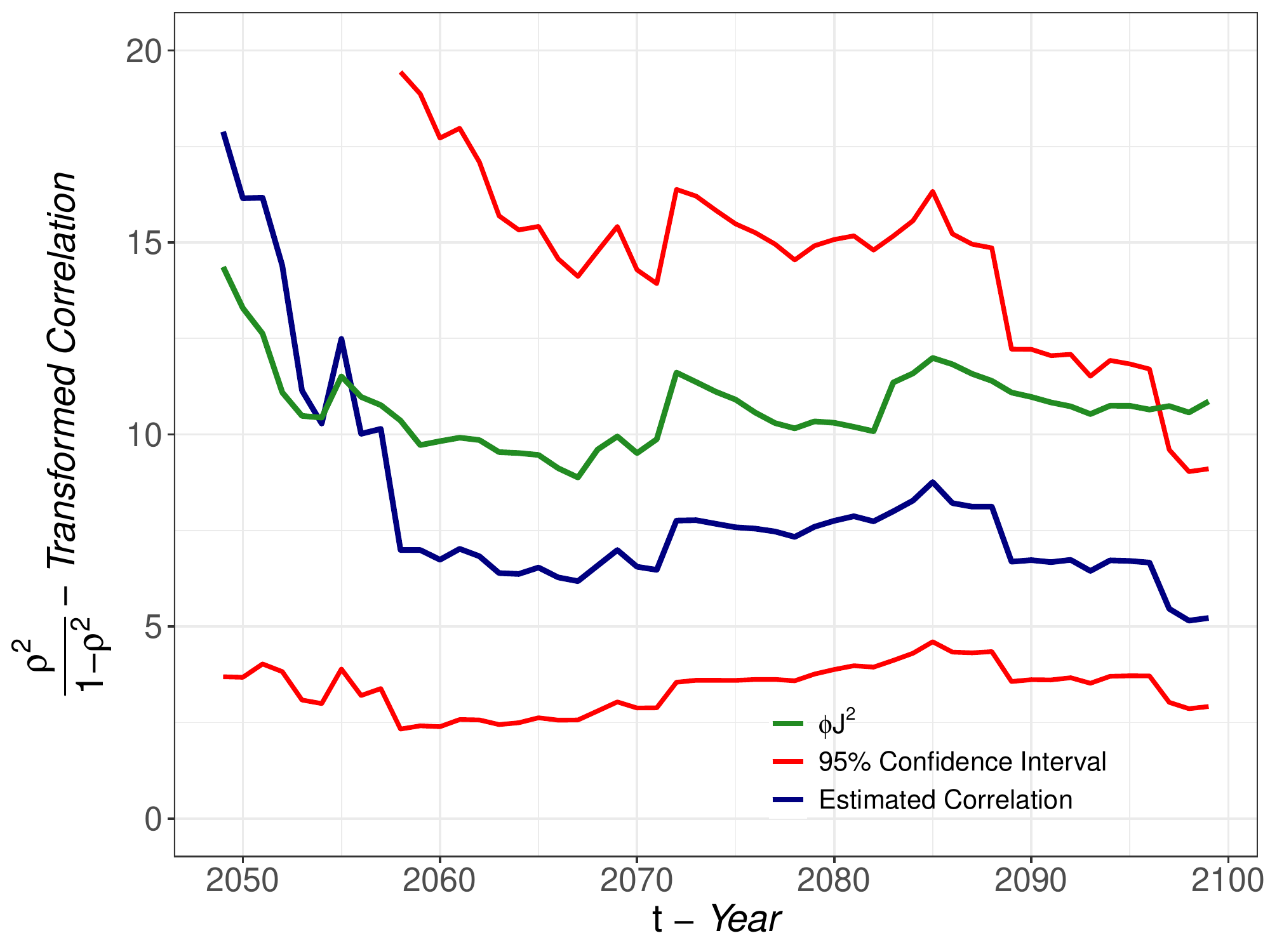}
    \caption{Long run transformed correlation against $J^2\phi$ for the NorESM2-LM data}
    \label{fig: J Eqn for Data}
\end{figure}

One notable result from Section \ref{Mathematical modelling} was the J Equation (Equation \ref{eqn: J Equation}), which presented the relationship between the correlation of frequency and aggregate risk and the dispersion statistic. Using the same methods as above, a long run average was calculated for the correlation as well as the dispersion statistic. The transformed correlation and corresponding confidence intervals was then plotted against the product of the long run values of $\phi$ and $J$, this is shown in Figure \ref{fig: J Eqn for Data}. 
\vspace{1em}

It must be noted that in Figure \ref{fig: J Eqn for Data} we do not have perfect equality, despite the green line of $\phi J^2$ mostly lying within the 95\% confidence interval. This is likely due to the sensitivity of correlation (as mentioned in Section \ref{Exploration of the relationship between Frequency and Aggregate Risk}). While these results held well, with $\approx3\%$ of the \say{$J^2\phi$} line lying outside of the 95\%  confidence interval, I decided to test the J-Equation further. Data was simulated over the same period, using the model shown below (\ref{Simulated GPD Model}) a time-varying Poisson process and Generalised Pareto Distribution (GPD) to generate counts respectively.

\begin{align} \label{Simulated GPD Model}
    N_t&\sim \text{Pois}(\alpha_0 +\alpha_1 t) &t=1,\dotsc ,60 \nonumber\\
    X_{it} &\sim \text{GPD}(0, \xi, \frac{1}{\beta_0 +\beta_1 t}) & i=1,\dotsc,N_t.
\end{align}

\begin{figure}[H]
    \centering
\includegraphics[height=0.425\textheight,keepaspectratio]{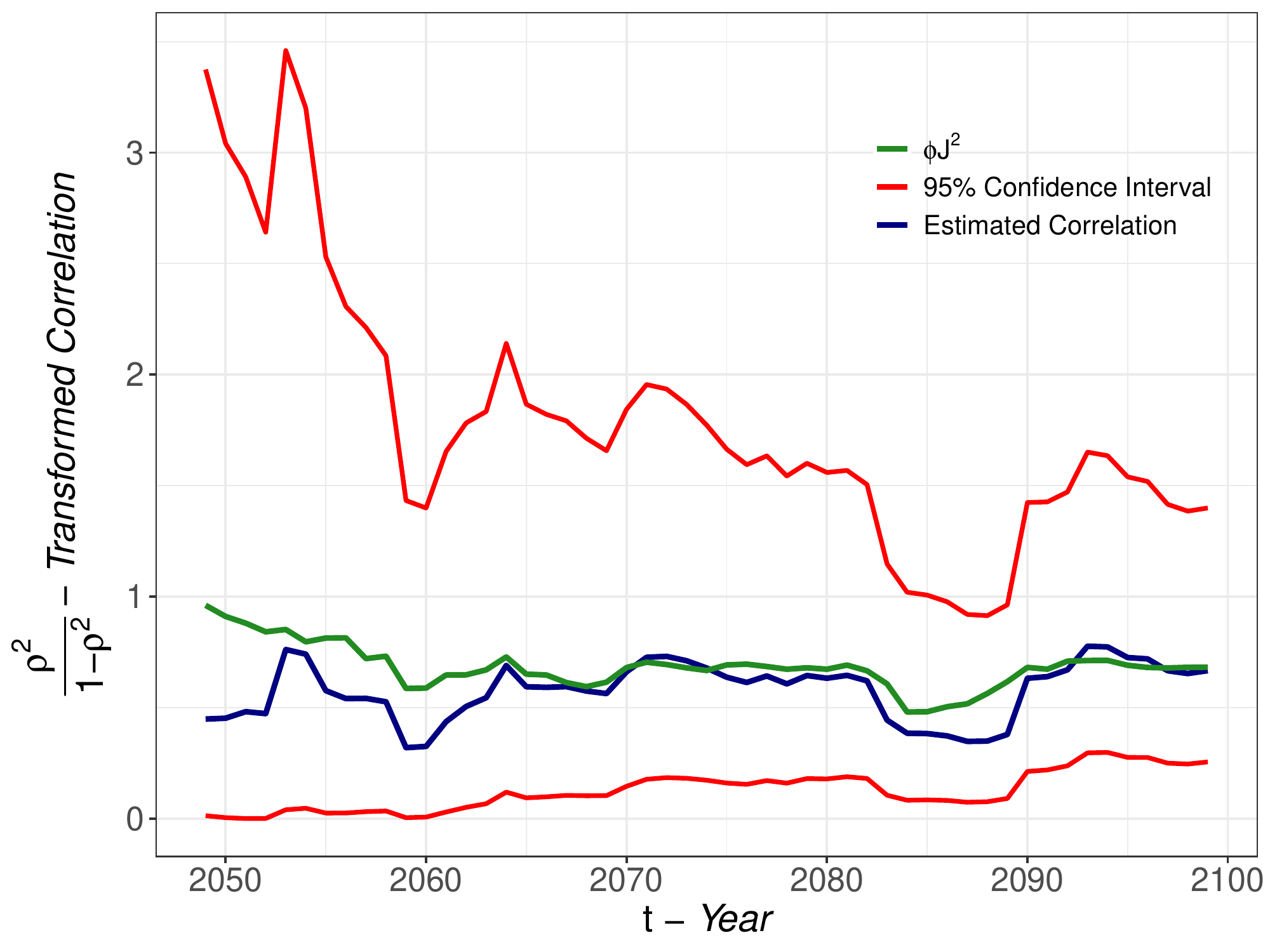}
    \caption{Long run transformed correlation against $J^2\phi$ for the simulated dataset}
    \label{fig: J Eqn for Simulated Data}
\end{figure}

As shown in Table \ref{table:1} the value of the ratio $J$ for the GPD is $J=1-2\xi$, as $\xi$ is fixed this value is invariant in time. As $N$ follows a Poisson distribution the value of the dispersion statistic is one so is also constant in time. Thus $J^2\phi=(1-2\xi)^2$, and as shown in Figure \ref{fig: J Eqn for Simulated Data}, the simulated data confirms the result more accurately, with both lines following roughly the same path.

\vspace{1em}

A likely reason for the NorESM2-LM dataset not following the J Equation is that the statistical assumptions I have made (for example independence between frequency and severity of storms) over-simplify the system - thus introducing error into the equation.

\subsection{Investigation of independence}\label{Investigation of Independence}

 Section \ref{Mathematical modelling} assumed that frequency and intensity of storms was conditionally independent. Independence can be quantified through correlation, events that are independent have correlation close, or equal to, zero. While the converse is not always true, investigation into the correlation between $N$ and $X$ gives a good measure of the accuracy of this assumption. A consequence of Wald's Equation is that the expected intensity is equal to the mean value of the aggregate risk. By computing this mean aggregate risk for each year, despite this being a rough approximation, I found no strong evidence of any relationship between frequency and intensity ($\rho = 0.1613$). 

\vspace{1em}

\begin{figure}[H]
    \centering
    \includegraphics[height=0.425\textheight,keepaspectratio]{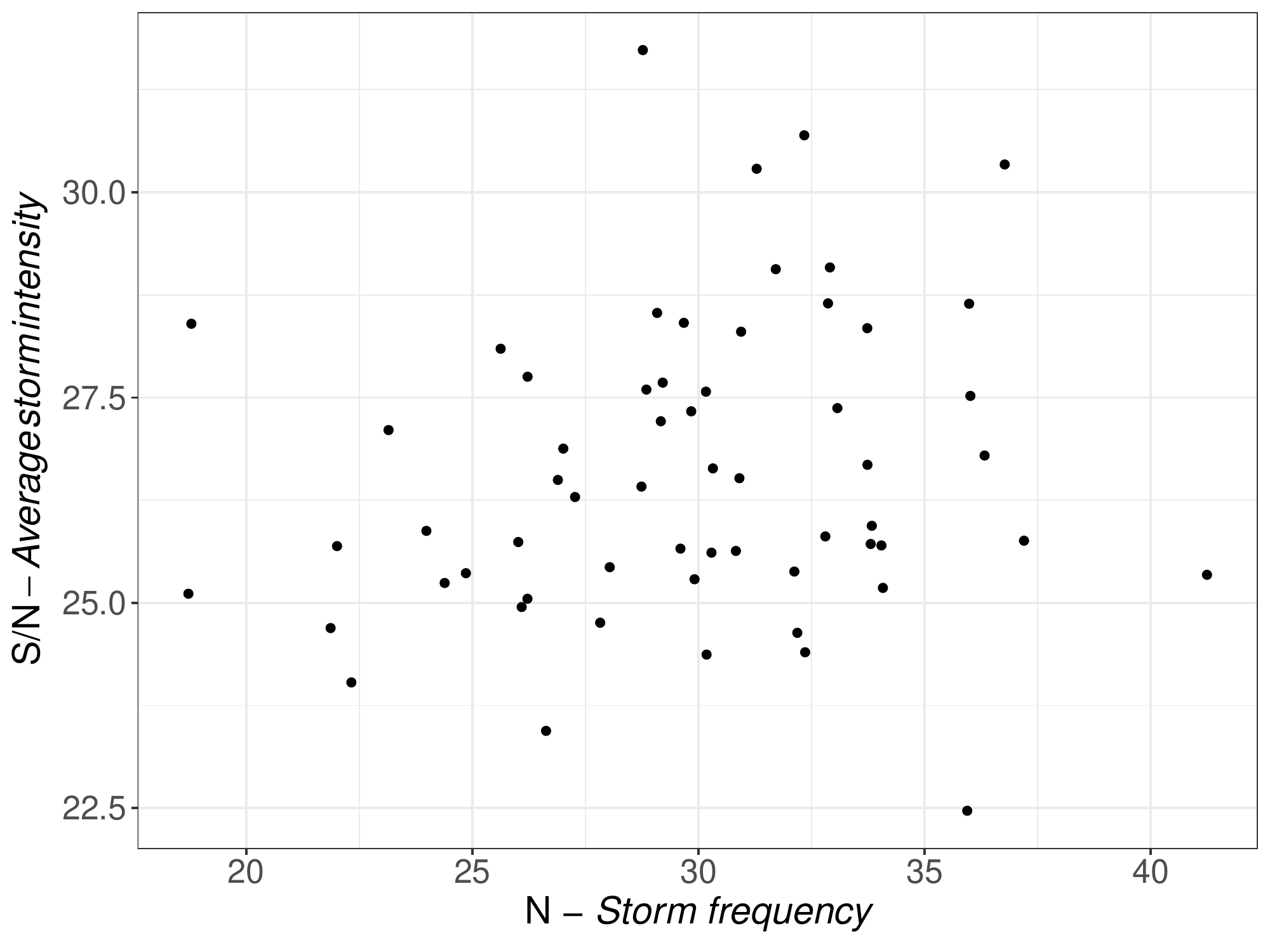}
    \caption{Scatter plot between the frequency $N$ and average intensity $\frac{S}{N}$ of storms in the dataset}
    \label{fig: Cor N X}
\end{figure}

 Figure \ref{fig: Cor N X} depicts the lack of overall correlation between frequency and intensity.  Any slight trend that exists may be as a result of both average intensity and frequency having dependence on time - not on each other. Given an overall near zero correlation, I concluded that the assumption of $N$ and $X$ being conditionally independent (when the year is fixed, they are independent) is coherent with the characteristics of the dataset.

\subsection{Exploration of the relationship between Frequency and {Aggregate Risk}}\label{Exploration of the relationship between Frequency and Aggregate Risk}

Section \ref{Mathematical modelling} showed that the correlation between frequency and aggregate risk (hereafter referred to as Cor$(N,S)$), was dependent on the dispersion statistic $\phi$ as well as the shape of $X$, the distribution of the intensities of cyclones. In this section I explore the value of Cor$(N,S)$, investigating how it changes in time and if the data fits with the theory presented.

\vspace{1em}

To explore this further I aimed to find the \say{long run} correlation, by computing the correlation  over an increasing number of years with the aim  of seeing convergence to a value. I computed approximate confidence intervals using the Fisher Transformation. This states that for \textit{i.i.d.}  pairs of random variables  $(X_i,Y_i)$ that follow a bivariate Normal distribution then the correlation may be approximated with a Normal distribution with mean $\frac{1}{2}\ln \frac{1+\rho}{1-\rho}$ and standard deviation $\frac{1}{\sqrt{N-3}}$. These confidence intervals are merely a guide, and should not be given too much importance, as the approximations of both $N$ and $S$ being following a Gaussian distribution as well as being identically distributed are a weak reflection of my true assumptions.

\begin{figure}[H]
    \centering
    \includegraphics[height=0.4\textheight,keepaspectratio]{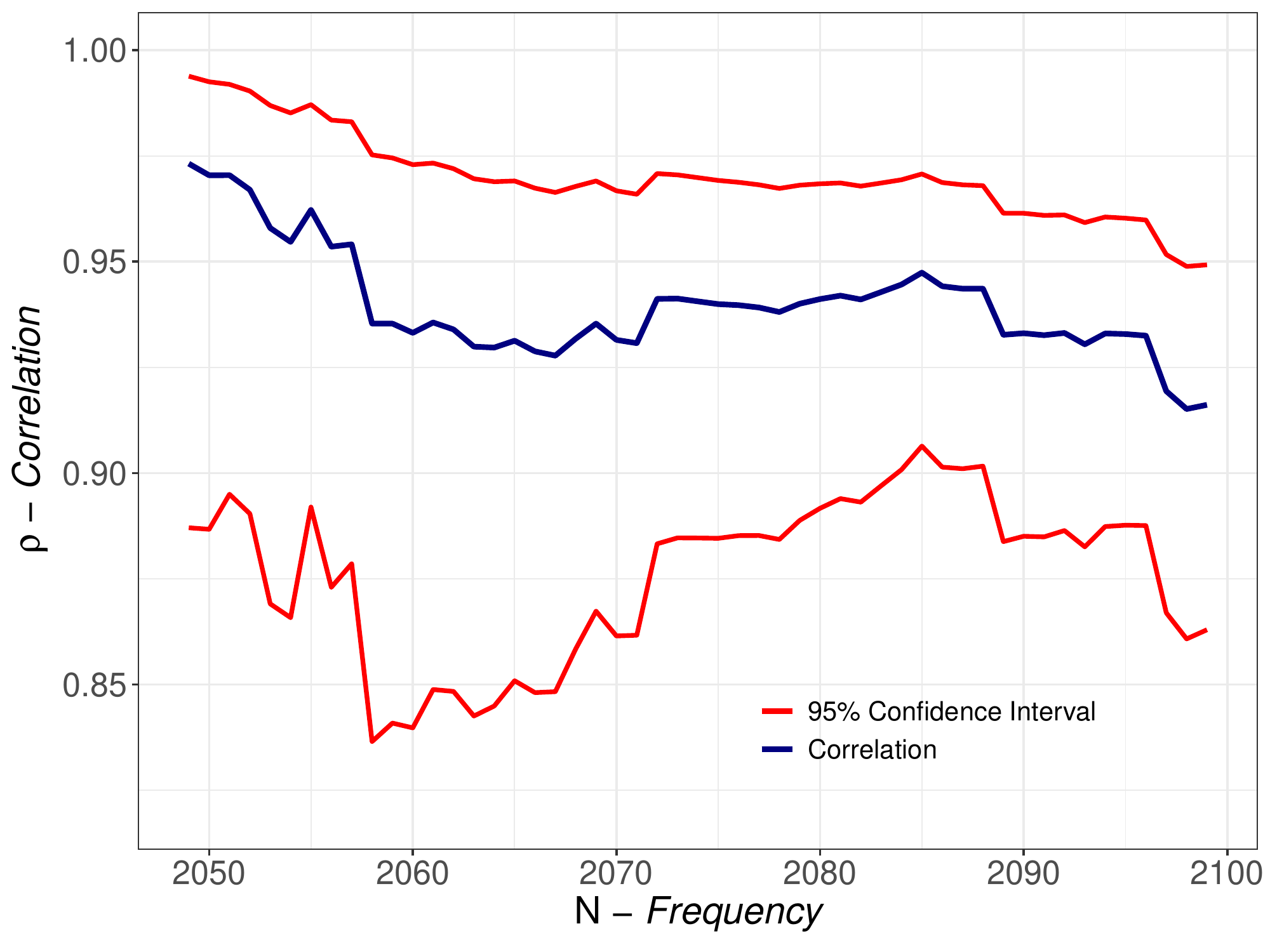}
    \caption{Moving average of correlation between frequency \& aggregate risk over the 6 decades of the dataset}
    \label{fig: Moving Window Correlation}
\end{figure}

Figure \ref{fig: Moving Window Correlation} provides a new perspective on the value of Cor$(N,S)$. The correlation appears to converged to an approximate value of $\rho \approx 0.93$ after around 20 years (the year 2060). The value of Cor$(N,S)$ appears to converge, with just slight \say{dip} in the final few years. These are years with anomalous storm intensities causing extremely high / low aggregate risk values compared to the number of events in that year.

\subsection{Analysis of Dispersion}\label{Analysis of Dispersion}

As noted by   \textcite{Economou}, clustering of storms is expected to increase in Europe in the future, and as concluded in Section \ref{Mathematical modelling}, the value of $\phi$ gives an insight into the characteristics of European storms. In this section, the aim is to explore the \say{long run} value of $\phi$ for the NorESM2-LM dataset, and discuss the consequences of the under-dispersion that is found.

\vspace{1em}

Using the formula for dispersion in Equation \ref{eqn: Dispersion Statistic Def} I was able to compute the \say{long-run} dispersion statistic to discover if the frequency of storms in the region follows a Poisson distribution (if $\phi=1$). This value is computed using the formula below:

\begin{align}\label{Long Run Dispersion Formulas}
\phi_t &=\frac{\text{Var}\big(N(t)\big)}{\text{E}\big[N(t)\big]} \\
\text{Where: }      \text{Var}\big[N(t)\big]&=\text{E}\big[N^2(t)\big]-\text{E}\big[N(t)\big]^2 \nonumber\\
&=\frac{\sum_{y=1}^{t} N_y^2 }{\sum_{y=1}^{t} 1 }- \left(\frac{\sum_{y=1}^{t} N_y }{\sum_{y=1}^{t} 1 }\right)^2\nonumber.
\end{align}

\begin{figure}[H]
    \centering
    \includegraphics[height=0.425\textheight,keepaspectratio]{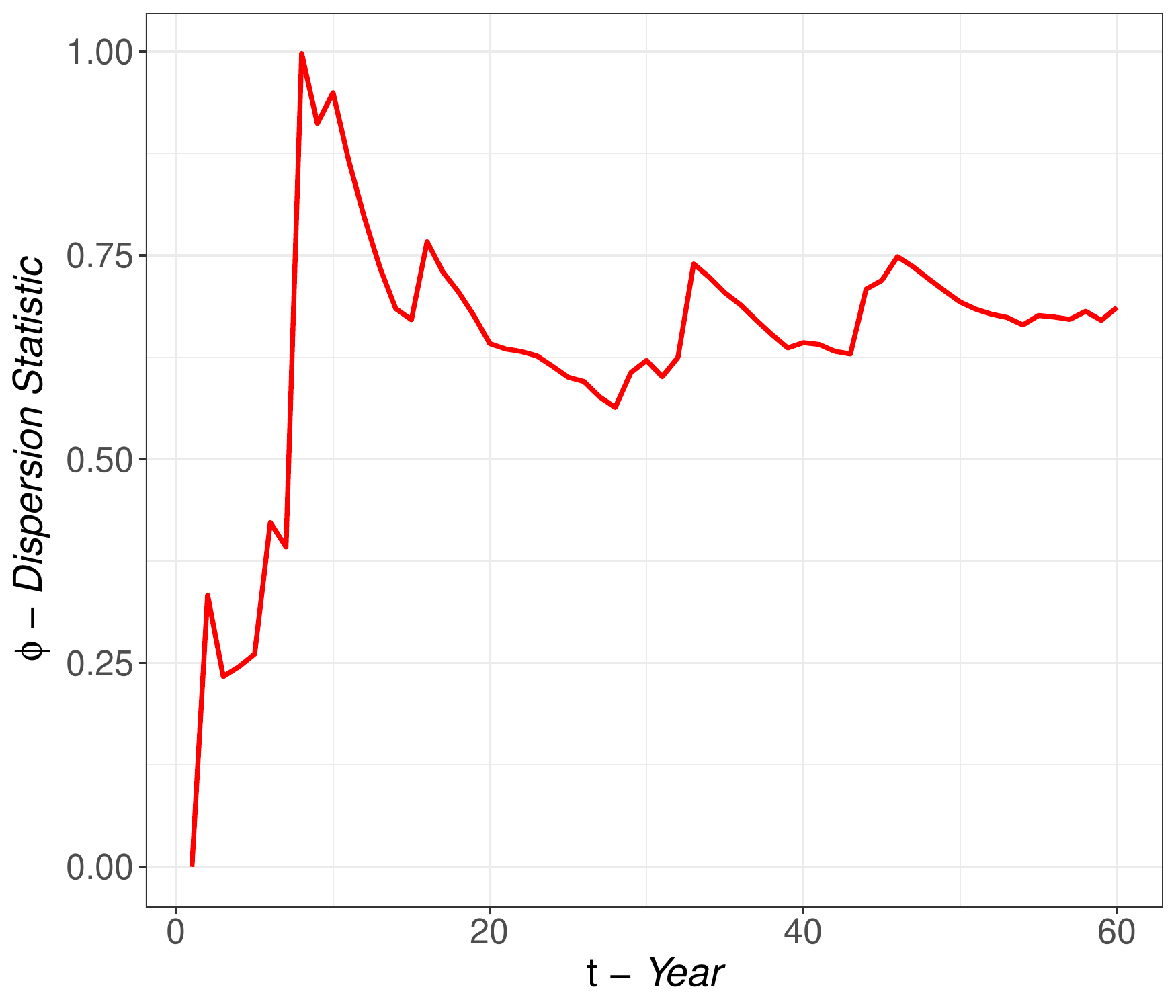}
    \caption{Long run transformed correlation against $J^2\phi$ for the simulated dataset}
    \label{fig: Long Run Dispersion Stat}
\end{figure}

As shown in Figure \ref{fig: Long Run Dispersion Stat}, we see that in fact storms in the region are under-dispersed, with a dispersion statistic converging to $\phi \approx 0.6662$, indicating more periodic storm occurrence. As Section \ref{Exploration of the relationship between Frequency and Aggregate Risk} concluded that the correlation between frequency and aggregate risk was fairly stable and time invariant, the implication that future storms will be more regular means aggregate risk could be predicted from the number of storms in that season.

\newpage

\section{Conclusion} \label{Conclusion}


In this project I have used random sums to investigate aggregate risk (Equation \ref{eqn: Aggregate Risk}), considering the frequency and intensity of storms as non-stationary in time. Furthermore, I have shown the link between aggregate risk and the historic equations from Wald (Equation \ref{eqn: Wald Eqn (E[S|T] Proof)}) and Blackwell and Girschick (Equation \ref{eqn: Variance of S Var(S|T)}). Not only does Wald's Equation hold theoretically, but I have proved it works for moving sample means. The new perspective of expected aggregate risk in a changing climate has implications on future research - a trend of increased storm activity would imply a rise in aggregate risk.

\vspace{1em}

While the relationship between the formula for aggregate risk and its' variance may not be so obvious, this equation allowed me to determine the covariance, and thus correlation, between storm frequency and aggregate risk. For a fixed year, I demonstrated that this covariance  was the product of the expected intensity of storms and the variance in their counts (Eqn. \ref{eqn: Covariance between N and S}). 
This also showed that the covariance between aggregate risk and the occurrence of storms is the product of the dispersion statistic and the expected value of the aggregate risk. If the dispersion statistic is stationary then the strength of the relationship between storm frequency and aggregate risk is proportional to the expected value of aggregate risk. 

\vspace{1em}

This equation for the covariance then allowed me to provide an expression for the correlation between the two variables. Alongside this, the introduction of the J ratio (the reciprocal of the coefficient of variation) as a measure of the shape of a distribution provided a unique approach to the correlation between frequency and aggregate risk. The J-Equation signifies a link between the clustering of storms and the strength of the dependency aggregate risk has on storm frequency.

\vspace{1em}

While the J-Equation did not  hold precisely with the dataset used, this is likely due to the size of the dataset and methods used to calculate correlation, which is quite volatile under the influence of outlying points in smaller datasets. For the simulated data, the J-Equation held, indicating that potentially some assumptions I have made introduce error. An example of this may be independence, between both storm intensity and frequency as well as the individual intensity of storms in a given year. While it is difficult to quantify independence in both of these cases, this is an extension to the project which would likely validate many of the results presented.

\vspace{1em}

A continuation of this project could investigate further non-stationary random sums and extreme values of aggregate risk, building upon the theory presented here to determine any trends and causes of extremes. Another extension would be to also undertake this analysis for the correlation between aggregate risk and storm intensity to see any historical changes, potentially providing insight into whether frequency or intensity is more influential in the value of aggregate risk. Further investigation of the J-Equation would be concerned with how the correlation between aggregate risk and storm frequency has changed over time, and how it is projected to increase under a changing climate (particularly under a variety of different data from CMIP6 models). One final continuation would be an investigation into whether the value of the J Ratio (the shape of the distribution of storm severity) has changed historically as well as how it may evolve under the  earth's changing climate.


\vspace{10em}
\seperate
\begin{center}
 \textbf{Acknowledgements}\\
   Recognition must go to my supervisor David Stephenson, without his expertise, support and guidance I could not have completed this project. I must also acknowledge Matthew Priestley for kindly supplying the CMIP6 dataset as well as drafts of his current work. Finally, thanks must also go to my parents and friends, whose words of encouragement helped me along the way.
   
\end{center}

\vspace{1em}

\seperate

\newpage
\section{Appendix} \label{Appendix}

\textbf{Derivation \ref{app: Covariance between X and S}:} Covariance between $X$ and $S$   (storm intensity and aggregate risk)

\begin{align}\label{app: Covariance between X and S}
 \text{cov}(X,S\,|\,t=T)&=\text{cov}(X_1,X_1 +\dots+ X_N\,|\,T,N) \nonumber\\
&=\text{cov}(X_1,X_1\,|\,T,N)+0+\dots+0 \nonumber\\
 &=\text{Var}(X\,|\,T).
\end{align}

\seperate

\textbf{Derivation \ref{app: Correlation between X and S}:} Correlation between $X$ and $S$   (storm intensity and aggregate risk)

\begin{align} \label{app: Correlation between X and S}
 \text{cor}(X,S\,|\,T)&=\frac{\text{cov}(X,S\,|\,T,N)}{\sqrt{\text{Var}(X\,|\,T)\text{Var}(S\,|\,T)}}\nonumber\\
&=\frac{\text{Var}(X\,|\,T)}{\sqrt{\text{Var}(X\,|\,T)\text{Var}(S\,|\,T)}}\nonumber\\
&=\sqrt{\frac{\text{Var}(X\,|\,T)}{\text{Var}(S\,|\,T)}}.
\end{align}

\seperate

{\large\textbf{Derivations of results in Table \ref{table:1}}}

\vspace{1em}

This section assumes, without proof, results for the expectation and variance of the  various distributions. We also assume the frequency of storms follows a non-stationary Poisson distribution with mean E$[N]=\text{Var}(N)=\lambda_T$ and define $\mu_T=\beta_0+\beta_1 T$.

\vspace{1em}

{\large \underline{The log-Normal Distribution}}

\vspace{1em}

If X follows a Logarithmic-Normal distribution then:
$$f(X; \mu_T, \sigma) = \frac{1}{x \sigma \sqrt{2 \pi}} \exp \left( -\frac{(\ln x - (\mu_T))^2}{2 \sigma^2} \right)$$

\newpage
The expectation and variance of these distributions are well-known and as such:
\begin{align*}
\text{E}[X\,|\,T]&= \exp \left( \mu_T + \frac{\sigma^2}{2} \right) , \\
\text{Var}(X\,|\,T)&= \left( e^{\sigma^2}-1\right) \exp \left(2\mu_T + \sigma^2 \right),\\
\text{E}[S\,|\,T]&= \lambda_T \exp \left(\mu_T + \frac{1}{2}\sigma^2 \right),\\
\text{Var}(S\,|\,T)&= \lambda_T  \exp \left( 2\mu_T + 2\sigma^2 \right).
\end{align*}

The correlation between frequency and aggregate risk is: 
\begin{align*}
     \text{cor}(N,S)&=\frac{\text{E}[X\,|\,T]}{\sqrt{\text{E}[X^2\,|\,T]}}\\
&= \frac{ \exp \left( \mu_T+\frac{1}{2}\sigma^2 \right)}{\sqrt{ \exp \left( 2\mu_T +2\sigma^2 \right) }}\\
     &= \exp \left( \mu_T +\frac{1}{2}\sigma^2-( \mu_T +\sigma^2) \right)\\
     &= e^{-\frac{1}{2}\sigma^2}.
\end{align*}

The value of $J^2$ is:
\begin{align*}
J^2&=\frac{\text{E}[X\,|\,T]^2}{\text{Var}(X]\,|\,T)}\\
&=  \frac{\exp \left( \mu + \frac{1}{2}\sigma^2 \right)  ^2}{\left( e^{\sigma^2}-1\right) \exp \left( 2\mu + \sigma^2 \right)}\\
 &=  \frac{\exp \left( 2\mu + \sigma^2 \right)}{\left( e^{\sigma^2}-1\right) \exp \left( 2\mu + \sigma^2 \right)}\\
     &=  \frac{1}{\left( e^{\sigma^2}-1\right)}.
\end{align*}

\newpage

{\large \underline{The Uniform Distribution}}

\vspace{1em}

If X follows a Uniform distribution then we assume that every storm intensity is equally likely, from zero to the maximum for that year, which we consider to be non-stationary in time.
\begin{align*}
X_{iT} &\sim \text{Unif}(0, \mu_T )\\
f(X_{iT}; \,\mu_T) &= \frac{1}{ \mu_T}
\end{align*}
If $Y \sim \text{Unif}(a,b)$ then:
\begin{itemize}
    \item E$[Y]=\frac{a+b}{2}$
    \item Var$(Y)=\frac{1}{12}(b-a)^2$
\end{itemize}

Thus:
\begin{align*} 
\text{E}[X\,|\,T]&=  \frac{1}{2}\mu_T,\\
\text{Var}(X \,|\, T) &= \frac{1}{12} \mu_T^2.
\end{align*}
The expected aggregate risk is:
\begin{align*} 
\text{E}[S\,|\,T]&=\text{E}[N\,|\,T]\text{E}[X_T\,|\,T]\nonumber\\
&=  \frac{1}{2}\lambda_T\mu_T.
\end{align*} 
The variance in aggregate risk is:
\begin{align*}
\text{Var}(S\, | \,T) &=  \text{E}[N\,|\,T] \text{E}[X^2 \, | \, T] \nonumber \\
 &=  \lambda_T \left[ \frac{1}{12} \mu_T^2 + \left( \frac{1}{2} \mu_T\right)^2 \right] \nonumber \\
 &= \frac{1}{3}\lambda_T \mu_T^2.
\end{align*}
The correlation between $N$ and $S$, if $X$ follows a Uniform distribution, is:
\begin{align*}
 \text{cor}(N,S) &=\frac{\text{E}[X]}{\sqrt{\text{E}[X^2]}}\\ 
&= \frac{\frac{1}{2}(\beta_0 +\beta_1 T)}{\sqrt{\frac{1}{3}(\beta_0 +\beta_1 T)^2}}\\
&=\frac{\sqrt{3}}{2}    .
\end{align*}
The value of $J^2:$
\begin{align*}
  J^2&=\frac{\text{E}[X\,|\,T]^2}{\text{Var}(X\,|\,T)}\\
  &=  \frac{(\frac{1}{2}a)^2}{\frac{1}{12}a^2}\\
  &=3.
  \end{align*}

\vspace{1em}

{\large \underline{The Gamma Distribution}}

\vspace{1em}

If $X$ follows a Gamma Distribution then:
\begin{align*}
X_{iT} &\sim \text{Gamma}\left(\theta, \frac{1}{\mu_T} \right)\\
f(X_{iT}; \,\mu_T, \theta) &=\frac{\mu_T^{-\theta}}{\Gamma (\theta)} x^{\theta-1} \exp \left(\frac{-x}{\mu_T}\right)
\end{align*}
\begin{center}
\textit{Where $\theta>0$ is the shape parameter.}
\end{center}

If $Y \sim \text{Gamma}(\alpha,\beta)$ then:
\begin{itemize}
    \item E$[Y]=\frac{\alpha}{\beta}$
    \item Var$(Y)=\frac{\alpha}{\beta^2}$
\end{itemize}

Therefore we see:
\begin{align*}
  \text{E}[X\,|\,T]&=\theta\mu_T,\\
     \text{Var}(X\,|\,T)&=\theta \mu_T ^2. 
\end{align*} 
The expected aggregate risk is:
\begin{align*}
  \text{E}[S\,|\,T]&=  \text{E}[N\,|\,T]  \text{E}[X_T\,|\,T]\\
&=  \lambda_T\theta \mu_T/
\end{align*} 
The variance of $S$ is:
\begin{align*} \label{gamma Var(S)}
\text{Var}(S\, | \,T)&=  \text{E}[N\,|\,T]  \text{E}[X^2\,|\,T]\\
&= \lambda_T \big(  \text{Var}(X\,|\,T)+ E[X \, | \, T]^2 \big) \\
 &= \lambda_T(\theta+\theta^2)\mu_T ^2. \\
\end{align*}

The correlation between frequency and intensity is:

\begin{align*}
  \text{cor}(N,S,|\,T)&=\frac{ \text{E}[X\,|\,T]}{\sqrt{ \text{E}[X^2\,|\,T]}}\\
&= \frac{\theta\mu_T}{\sqrt{ (\theta+\theta^2 )\mu_T^2  }}\nonumber\\
    &= \frac{\theta}{\sqrt{ \theta+\theta^2 }}= \frac{1}{\sqrt{ \frac{1}{\theta}+1}}.
\end{align*}

The value of $J^2$ is:
\begin{align*}
    J^2&=\frac{\text{E}[X\,|\,T]^2}{\text{Var}(X\,|\,T)}\\
    &=\frac{(\theta \mu_T)^2}{\theta\mu_T^2}\\
    &=\theta.
\end{align*}


\newpage

{\large \underline{The Exponential Distribution}}

\vspace{1em}

It is important to note that the Exponential distribution is simply the Gamma distribution, but with  $\theta=1$, thus all results can be easily obtained.

For: \begin{align*}
X_{iT} &\sim \text{Exp}\left(\frac{1}{\mu_T} \right)\\
f(X_{iT}; \,\mu_T) &=\frac{1}{\mu_T} \exp \left(\frac{-x}{\mu_T}\right)
\end{align*}
We have:
\begin{itemize}
    \item $\text{E}[S\,|\,T] =\lambda_T\theta \mu_T=\lambda_T \mu_T$
    \item $\text{Var}(S\, | \,T)=\lambda_T(\theta+\theta^2)\mu_T ^2=2\lambda_T\mu_T ^2 $
    \item $\text{cor}(N,S,|\,T)= \frac{\theta}{\sqrt{ \theta+\theta^2 }}=\frac{\sqrt{2}}{2}$
    \item $J^2=\theta=1$
\end{itemize}


{\large \underline{The Generalised-Pareto Distribution (GPD)}}

\vspace{1em}

If X follows a GPD then:
$$X \sim \text{GPD}\left(\kappa,\frac{1}{\mu_T}, \xi\right) $$
\begin{gather*}
f\left(X;\kappa,\frac{1}{\mu_T}, \xi\right) =
\mu_T\bigg(1+\xi\mu_T (x-\kappa)\bigg)^{\left(-\frac{1}{\xi} -1\right)} \\
\end{gather*}
However often only extreme storms are considered, and as such we can \say{centre} the distribution around the threshold we want to consider, this means we can set $\kappa=0$. If $Y \sim \text{GPD}(0,\sigma,\xi)$ then:
\begin{itemize}
    \item $\text{E}[Y]=\frac{\sigma}{1-\xi}$ when $\xi< 1$
    \item $\text{Var}(Y)=\frac{\sigma^2}{(1-\xi)^2 (1-2\xi)}$ when $\xi< \frac{1}{2}$
\end{itemize}
As a result, we have:
\begin{align*}
    \text{E}[X\,|\,T]&=\frac{1}{\mu_T(1-\xi)},\\
    \text{Var}(X\,|\,T)&=\frac{1}{\mu_T^2(1-\xi)^2 (1-2\xi)}.
\end{align*}

The expected value of the yearly aggregate risk is thus:
\begin{align*}
\text{E}[S\,|\,T]&=\text{E}[N\,|\,T] E [X_{T}\,|\,T]\\
&=\frac{\lambda_T}{\mu_T(1-\xi)}.
\end{align*} 

The variance in aggregate risk is:
\begin{align*}
 \text{Var}(S\, | \,T) &=\text{E}[N\,|\,T]\text{E}[X^2\,|\,T]\\
&=\lambda_T \big(\text{Var}(X\,|\,T)+\text{E}[\text{E}\,|\,T]^2 \big)\\
 &=\frac{\lambda_T}{\mu_T^2(1-\xi)^2(1-2\xi)}+\lambda_T\bigg(\frac{1}{\mu_T(1-\xi)}\bigg)^2  \\
&=\frac{2\lambda_T}{\mu_T^2(1-\xi)(1-2\xi)}.
\end{align*}

The correlation between aggregate risk and storm frequency is:
\begin{align*}
\text{cor}(N,S)&=\frac{\text{E}[X\,|\,T]}{\sqrt{\text{E}[X^2\,|\,T]}}\\
&=\frac{1}{\mu_T(1-\xi)}\sqrt{\frac{\mu_T^2(1-\xi)(1-2\xi)}{2}}\\
&=\sqrt{\frac{1-2\xi}{2-2\xi}}.
\end{align*}
\newpage
The value of $J^2$ when $X$ follows a GPD is:

\begin{align*}
J^2&=\frac{\text{E}[X\,|\,T]^2}{\text{Var}(X\,|\,T)}\\
&=(\frac{\sigma}{1-\xi})^2\frac{(1-\xi)^2 (1-2\xi)}{\sigma^2}\\
&= 1-2\xi.
\end{align*}

\seperate

\textbf{Proof \ref{App: Long Run Wald Derivation}} - Proof that Wald's Equation (Equation \ref{eqn: Wald Eqn (E[S|T] Proof)}) holds for long run expectations as shown in Equations \ref{N S X Equations} provided $N_y \ne 0$.

\begin{align}\label{App: Long Run Wald Derivation}
     \text{E}[N(t)]\text{E}[X(t)]&=\nonumber\\
     \lim_{t \to \infty}\frac{\sum_{y=1}^{t} N_y }{\sum_{y=1}^{t} 1 }\cdot \frac{\sum_{y=1}^{t} \sum_{i=1}^{N_y}X_{yi}} {\sum_{y=1}^{t} \sum_{i=1}^{N_y} 1 }&=\nonumber\\
    \lim_{t \to \infty} \frac{\sum_{y=1}^{t} N_y }{\sum_{y=1}^{t} \sum_{i=1}^{N_y} 1 }\cdot \frac{\sum_{y=1}^{t} \sum_{i=1}^{N_y}X_{yi}}{\sum_{y=1}^{t} 1 }&=\nonumber\\
    \lim_{t \to \infty} \frac{\sum_{y=1}^{t} N_y }{\sum_{y=1}^{t} {N_y} }\cdot   \frac{\sum_{y=1}^{t} \sum_{i=1}^{N_y}X_{yi}}{\sum_{y=1}^{t} 1 }&=\nonumber\\
     \lim_{t \to \infty} \frac{\sum_{y=1}^{t} N_y }{\sum_{y=1}^{t} \sum_{i=1}^{N_y} 1 } &= \text{E}[S(t)]\nonumber\\
   \therefore \text{ as } t \to \infty \hspace{5pt} \text{E}[N(t)]\text{E}[X(t)]&=\text{E}[S(t)].
 \end{align}
 
\seperate

\begin{figure}[H]
    \centering
    \includegraphics[width=0.7\textwidth,height=\textheight,keepaspectratio]{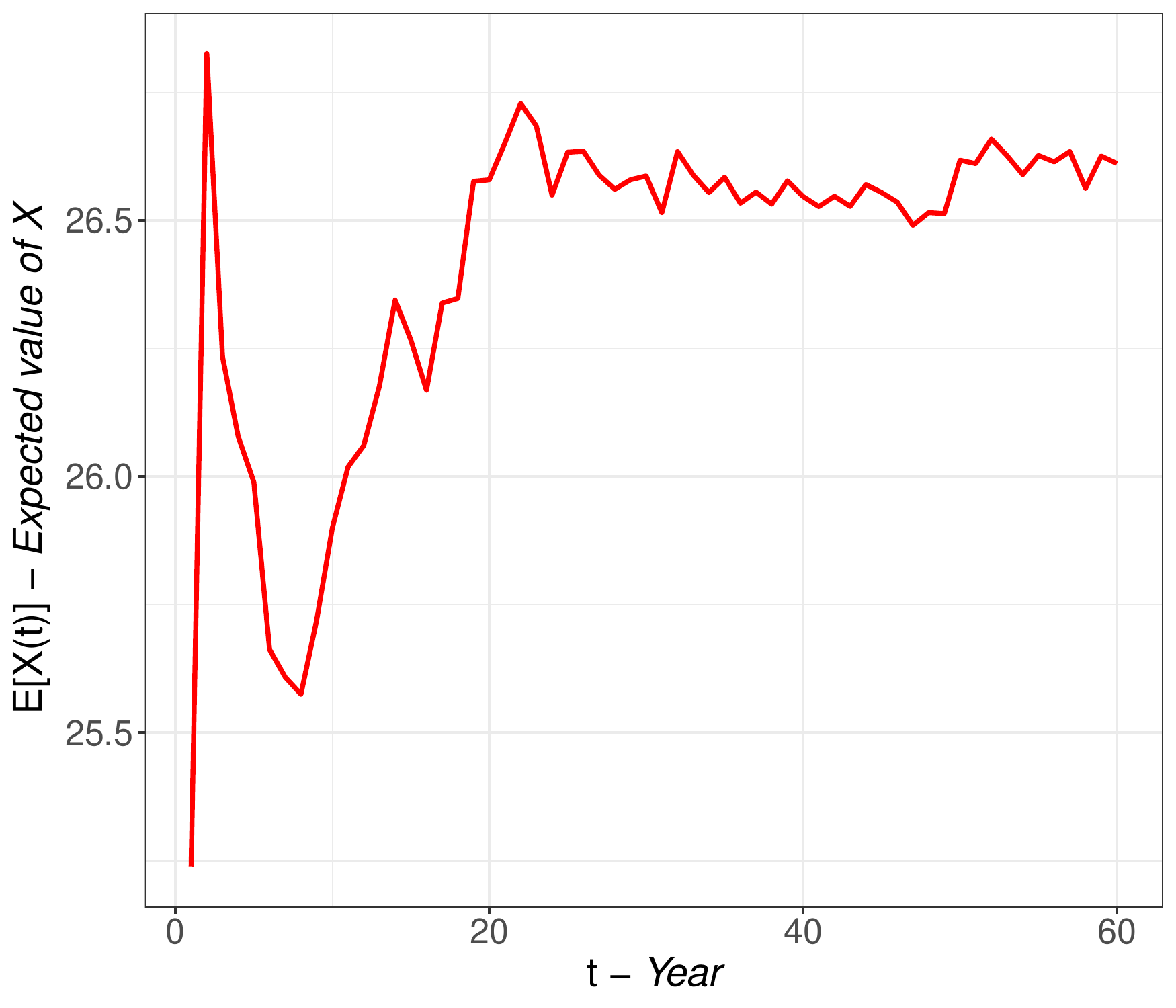}
    \caption{Long run average of $X(t)$ over the duration of the dataset.}
    \label{App: Long Run X(t)}
\end{figure}

\newpage
\printbibliography

@online{air-alert,
    organization = "AIR Worldwide",
    title = "Winter Storm Ciara Sabine",
    year = "2020",
    url  = "https://alert.air-worldwide.com/EventSummary.aspx?e=938&tp=31",
    addendum = "(accessed: 08.03.2021)",
}

@article{hunter, 
    author = "A. Hunter and D.B. Stephenson and T. Economou and M. Holland and I. Cook",
    title = "{New perspectives on the collective risk of extratropical cyclones}",
    journal = "Q. J. R. Meteorol Soc.",
    volume = "142",
    number = "694",
    pages = "243--256",
    year = "2015",
    DOI = "https://doi.org/10.1002/qj.2649"
    
}

@article{mailier,   
    author = {P.J. Mailier and D.B. Stephenson and C.A.T. Ferro},
    title = {Serial Clustering of Extratropical Cyclones},
    journal = "Mon. Wea. Rev. American Meteorol. Soc.",
    volume = "134",
    number = "8",
    pages = "2224--2240",
    year = "2006",
    DOI = "https://doi.org/10.1175/MWR3160.1"
}

@phdthesis{hunter_thesis,  
    author = {A. Hunter},
    title = {Quantifying and understanding the aggregate risk of natural hazards},
    institution = "University of Exeter",
    location = "Exeter,UK",
    library = "Retrieved from EThOS Database",
    year = "2014",
    DOI = "https://ethos.bl.uk/OrderDetails.do?did=1&uin=uk.bl.ethos.630861"
}

@article{steptoe,  
    author = {H. Steptoe and S.E.O. Jones and H. Fox},
    title = {Correlations Between Extreme Atmospheric Hazards
and Global Teleconnections: Implications
for Multihazard Resilience},
    journal = "Review of Geophysics",
    volume = "56",
    number = "1",
    pages = "50--78",
    year = "2018",
    DOI = "https://doi.org/10.1002/2017RG000567"
}

@article{cusack,  
    author = {S. Cusack},
    title = {The observed clustering of damaging extratropical cyclones in Europe},
    journal = " Natural Hazards and Earth System Sciences Discussions",
    volume = "16",
    number = "4",
    pages = "901--913",
    year = "2016",
    DOI = "https://doi.org/10.5194/nhess-16-901-2016"
}

@article{walz2,   
    author = {M.A. Walz and D.J. Befort and N.O. Kirchner-Bossi and U. Ulbrich and G.C. Leckebusch},
    title = {Modelling serial clustering and inter‐annual variability of European winter windstorms based on large‐scale drivers},
    journal = "International Journal of Climatology",
    volume = "38",
    number = "7",
    pages = "3044--3057",
    year = "2018",
    DOI = "https://doi.org/10.1002/joc.5481"
}

@article{priestley,
    author = {M.D.K. Priestley and H.F. Dacre and L.C. Shaffrey and K.I. Hodges and J.G. Pinto},
    title = {The role of European windstorm clustering for extreme seasonal losses as determined from a high resolution climate model},
    journal = "Natural Hazards and Earth System Sciences",
    volume = "18",
    number = "11",
    pages = "2991--3006",
    year = "2018",
    DOI = "https://doi.org/10.5194/nhess-18-2991-2018"
}

@article{Economou,
    author = {T. Economou and D.B. Stephenson and J.G. Pinto and L.C. Shaffrey and G. Zappa},
    title = {Serial clustering of extratropical cyclones in a multi-model ensemble of historical and future simulations},
    journal = "Quarterly Journal of the Royal Meteorological Society",
    volume = "141",
    number = "693",
    pages = "3076--3087",
    year = "2015",
    DOI = "https://doi.org/10.1002/qj.2591"
}

@article{Raschke,  
AUTHOR = {M. Raschke},
TITLE = {Statistical detection and modeling of the over-dispersion of winter storm occurrence},
JOURNAL = {Natural Hazards and Earth System Sciences},
VOLUME = {15},
number = {8},
YEAR = {2015},
PAGES = {1757--1761},
DOI = {https://doi.org/10.5194/nhess-15-1757-2015}}

@article{wald,  
    author = {A. Wald},
    title = {Sequential Tests of Statistical Hypotheses},
    journal = "Annals of Mathematical Statistics",
    volume = "16",
    number = "2",
    pages = "117--186",
    year = "1945",
    DOI = "https://doi.org/10.1214/aoms/1177731118"
}

@article{BandG,  
    author = {D. Blackwell  and   M.A. Girshick},
    title = {A Lower Bound for the Variance of Some Unbiased Sequential Estimates},
    journal = "Annals of Mathematical Statistics",
    volume = "18",
    number = "2",
    pages = "277--280",
    year = "1947",
    DOI = "https://doi.org/10.1214/aoms/1177730444"
}

@article{priestleydata,
    author = {M.D.K. Priestley and D. Ackerley and J.L. Catto and K.I. Hodges and R.E. McDonald and R.W. Lee },
    title = {An Overview of the Extratropical Storm Tracks in CMIP6 Historical Simulations},
    journal = "Journal of Climate",
    volume = "33",
    number = "15",
    pages = "6315--6343",
    year = "2020",
    DOI = "https://doi.org/10.1175/JCLI-D-19-0928.1"
}

@article{priestly2021,
author={M.D.K. Priestley and J.L. Catto},
title={Future changes in the extratropical storm tracks and associated
cyclone circulations},
pubstate={\textit{to be submitted}},
date={2021}}

@article{Kolmogorov,
    author = {A.N. Kolmogorov},
    title = {Sur la loi forte des grandes nombres},
    journal = "Acad. Sci. Paris",
    volume = "191",
    pages = "910--912",
    year = "1930",
    URL = "https://gallica.bnf.fr/ark:/12148/bpt6k31445"
}

@article{zappa,
    author = {G. Zappa and L.C. Shaffrey and K.I. Hodges and P.G. Sansom and D.B. Stephenson},
    title = {  A Multimodel Assessment of Future Projections of North Atlantic and European Extratropical Cyclones in the CMIP5 Climate Models },
    journal = "Journal of Climate",
    volume = "26",
    number = "16",
    pages = "5846--5862",
    year = "2013",
    DOI = "https://doi.org/10.1175/JCLI-D-12-00573.1"
}

@Article{cmip6_pathways,
AUTHOR = {O'Neill, B. C. and Tebaldi, C. and van Vuuren, D. P. and Eyring, V. and Friedlingstein, P. and Hurtt, G. and Knutti, R. and Kriegler, E. and Lamarque, J.-F. and Lowe, J. and Meehl, G. A. and Moss, R. and Riahi, K. and Sanderson, B. M.},
TITLE = {The Scenario Model Intercomparison Project (ScenarioMIP) for CMIP6},
JOURNAL = {Geoscientific Model Development},
VOLUME = {9},
YEAR = {2016},
NUMBER = {9},
PAGES = {3461--3482},
URL = {https://gmd.copernicus.org/articles/9/3461/2016/},
DOI = {https://doi.org/10.5194/gmd-9-3461-2016}
}

\end{document}